\documentclass[12pt]{article}
\usepackage{amsmath,amsfonts,amssymb}
\usepackage{stmaryrd}
\usepackage{lscape}
\usepackage{bm}
\usepackage{graphicx}
\usepackage[left=2cm,right=2cm,top=3cm,bottom=3cm]{geometry}
\usepackage{lscape}
\usepackage{url}
\newcommand{\rref}[1]{(\ref{#1})}
\newcommand{\p}{\partial}
\newcommand{\NR}{\mathbb{R}}

\newcommand{\bes}{\begin{equation}\begin{split}}
\newcommand{\ees}{\end{split}\end{equation}}
\newcommand{\sine}[1]{\sin(#1)}
\newcommand{\cosine}[1]{\cos(#1)}
\newcommand{\pd}[2]{\frac{\p #1}{\p #2}}

\newcommand{\bu}{\bar{u}}
\newcommand{\bx}{\bar{x}}

\newcommand{\bfx}{{\bf x}}
\newcommand{\bfu}{{\bf u}}
\DeclareMathOperator{\sech}{sech}
\def\tr{{\mathrm{Tr}}}
\numberwithin{equation}{section}
\newtheorem{prop}{Proposition}

\title{Conditionally invariant solutions of the rotating shallow water wave equations}
\author{Benoit Huard\\ 
D{\'e}partement de math{\'e}matiques et de statistique,\\
C.P. 6128, Succc.\ Centre-ville, Montr{\'e}al, (QC) H3C 3J7, Canada}

\begin{document}

\maketitle

\begin{abstract}
This paper is devoted to the extension of the recently proposed conditional symmetry method to first order nonhomogeneous quasilinear systems which are equivalent to homogeneous systems through a locally invertible point transformation. We perform a systematic analysis of the rank-$1$ and rank-$2$ solutions admitted by the shallow water wave equations in $(2+1)$ dimensions and construct the corresponding solutions of the rotating shallow water wave equations.  These solutions involve in general arbitrary functions depending on Riemann invariants, which allow us to construct new interesting classes of solutions.
\end{abstract}

\section{Introduction}
In this paper, we use the conditional symmetry method in the context of Riemann invariants ({\bf CSM}) as presented in \cite{Grundland07-1} to obtain conditionally invariant solutions of the rotating shallow water wave ({\bf RSWW}) equations with a flat bottom topography
\begin{equation}
\label{RSWW-eqs}
\begin{cases}
& u_t + u u_x + v u_y + g h_x = 2 \Omega v, \quad \Omega\in \mathbb{R},\\
& v_t + u v_x + v v_y + g h_y = -2 \Omega u,\\
& h_t + u h_x + v h_y + h(u_x + v_y) = 0,\\
\end{cases}
\end{equation}
where we denote by ${\bf x} = (t,x,y)$ and ${\bf u} = (u,v,h)$ the independent and dependent variables respectively.  Here, $u$ and $v$ stand for the velocity vector fields, $h$ represents the height of the fluid layer,$g$ is the gravitational constant and $\Omega$ characterizes the constant angular velocity of the fluid around the $z$-axis induced by a Coriolis force. It can be proved using the chain rule, see \cite{Chesnokov-RSWW}, that if a set of functions $u'(t',x',y'), v'(t',x',y'), h'(t',x',y')$ satisfies the irrotational shallow water wave equations ({\bf SWW})
\begin{equation}
\label{SWW-eqs}
\Delta'({\bf x},{\bf u}) : \begin{cases}
&{u'}_{t'} + u' {u'}_{x'} + v' {u'}_{y'} + g {h'}_{x'} = 0,\\
&{v'}_{t'} + u' {v'}_{x'} + v' {v'}_{y'} + g {h'}_{y'} = 0,\\
&{h'}_{t'} + u' {h'}_{x'} + v' {h'}_{y'} + {h'} ({u'}_{x'} + {v'}_{y'}) = 0,
\end{cases}
\end{equation}
then the functions $u(t,x,y), v(t,x,y), h(t,x,y)$ defined by
\begin{equation}
\label{SWW-RSWW-iso}
\begin{split}
&t' = -\frac{1}{2\Omega} \cot{\left(\Omega t\right)}, \quad x' = \frac{1}{2} \left(y-x \cot{\left(\Omega t\right)}\right),\quad y' = -\frac{1}{2} \left(x+y \cot{\left(\Omega t\right)}\right),\\
&u' = -\frac{1}{2} \big(u \sin{(2\Omega t)} - v (1- \cos{(2\Omega t)}) - 2\Omega x\big), \\
&v' = -\frac{1}{2} \big(u (1-\cos{(2\Omega t)}) + v \sin{(2\Omega t)} - 2\Omega y\big), \quad h' = \frac{h}{2} (1-\cos{(2\Omega t)}),
\end{split}
\end{equation}
form a solution of the RSWW equations.  

The task of constructing invariant solutions of systems (\ref{RSWW-eqs}) and (\ref{SWW-eqs}) using the classical Lie approach was undertaken by several authors.  A systematic classification of the subalgebras of the symmetry algebra of the equations describing a rotating shallow water flow in a rigid ellipsoidal bassin was performed in \cite{Levi-Winternitz-Nucci-Rogers-1989} and many invariant solutions were obtained.  In \cite{Chesnokov-RSWW}, the author introduced the transformation (\ref{SWW-RSWW-iso}) to generate invariant solutions of (\ref{RSWW-eqs}) from known invariant solutions of the homogeneous system (\ref{SWW-eqs}), previously computed in \cite{Chesnokov-SWW} .

The CSM approach to be used in this paper was developed progressively and applied in \cite{Conte-Grundland-Huard-JPhysA-2009,Grundland07-1,Grundland-Huard-JNMP} in order to construct rank-2 and rank-3 solutions to the equations governing the flow of an isentropic fluid.  The main feature of this approach, which proved to be less restrictive than the generalized method of characteristics \cite{Grundland07-1}, is that the obtained rank-$k$ solutions can depend on many arbitrary functions of many independent variables, called Riemann invariants.  Through a judicious selection of these arbitrary functions, it is possible to construct solutions of the considered homogeneous system which are bounded everywhere, even when the Riemann invariants admit a gradient catastrophe \cite{Conte-Grundland-Huard-JPhysA-2009}.  Although the applicability of the CSM approach is technically restricted to first order homogenous hyperbolic quasilinear systems, the objective of the present paper is to apply it to the RSWW equations (\ref{RSWW-eqs}) through the transformation (\ref{SWW-RSWW-iso}).  Large classes of implicit rank-$k$ solutions are then constructed for the SWW and RSWW equations, including bumps, kinks and periodic solutions.  

The paper is organized as follows.  We give in Section 2 the symmetry algebra of system (\ref{RSWW-eqs}) and construct the point transformation (\ref{SWW-RSWW-iso}) relating systems (\ref{RSWW-eqs}) and (\ref{SWW-eqs}). Section 3 contains a brief review of the conditional symmetry method in the context of Riemann invariants for homogeneous systems and we present many interesting rank-1 and rank-2 solutions to the SWW-equations (\ref{SWW-eqs}) together with corresponding solutions to the RSWW equations (\ref{RSWW-eqs}).  Results and perspectives are summarized in Section 4.


\section{The symmetry algebra}
The classical Lie symmetry algebra admitted by system (\ref{RSWW-eqs}) is generated by vector fields of the form
\begin{equation}
\label{symmetry-generator}
X = \xi^1 ({\bf x}, {\bf u}) \p_t +  \xi^2 ({\bf x}, {\bf u}) \p_x +  \xi^3 ({\bf x}, {\bf u}) \p_y +  \eta^1 ({\bf x}, {\bf u}) \p_u + \eta^2 ({\bf x}, {\bf u}) \p_v + \eta^3 ({\bf x}, {\bf u}) \p_h.
\end{equation}
The requirement that the generator \rref{symmetry-generator} leave system \rref{RSWW-eqs} invariant yields an overdetermined system of linear equations for the functions $\xi^i ({\bf x},{\bf u})$ and $\eta^i ({\bf x},{\bf u})$, $i=1,2,3$ \cite{Olver-2000}.  Since this step is completely algorithmic and involves tidy computations, many computer programs have been designed to derive these determining equations, see \cite{Hereman-review-94} for a complete review.  The package {\it symmgrp2009.max} \cite{Hereman-symmgrp,Hereman-symmgrp2009} for the computer algebra system {\it Maxima} has been used in this work to obtain the determining equations of the RSWW equations \rref{RSWW-eqs} and solve them partially in a recursive way.  
Solving them shows that the Lie algebra $\mathcal{L}$ of point symmetries of the RSWW equations \rref{RSWW-eqs} is nine-dimensional and is generated by the following differential generators
\begin{equation}
\label{Algebra-RSWW}
\begin{split}
&P_0 = \p_t, \quad P_1 = \p_x, \quad P_2 = \p_y, \quad L = y\p_x - x \p_y + v\p_u - u \p_v,\\
&G_1 = -\frac{1}{2\Omega} \cosine{2\Omega t} \p_x + \frac{1}{2\Omega} \sine{2\Omega t} \p_y + \sine{2\Omega t} \p_u + \cosine{2\Omega t} \p_v,\\
&G_2 = \frac{1}{2\Omega} \sine{2\Omega t} \p_x + \frac{1}{2\Omega} \cosine{2\Omega t} \p_y + \cosine{2\Omega t} \p_u - \sine{2\Omega t} \p_v,\\
&D = x\p_x + y\p_y + u\p_u + v \p_v + 2h\p_h,\\
&Z_1 = \sine{2\Omega t}\p_t + \Omega\left[x\cosine{2\Omega t} + y\sine{2\Omega t}\right]\p_x + \Omega\left[y\cosine{2\Omega t}-x\sine{2\Omega t}\right] \p_y \\
& \qquad + \Omega\left[(2\Omega y - u)\cosine{2\Omega t}  - (2\Omega x - v)\sine{2\Omega t} \right]\p_u  \\
& \qquad - \Omega\left[(2\Omega x + v)\cosine{2\Omega t} + (2\Omega y + u)\sine{2\Omega t}\right] \p_v - 2 \Omega h \cosine{2\Omega t} \p_h,\\
& Z_2 = \cosine{2\Omega t}\p_t + \Omega\left[y\cosine{2\Omega t} - x\sine{2\Omega t}\right]\p_x - \Omega\left[x\cosine{2\Omega t}+y\sine{2\Omega t}\right] \p_y \\
& \qquad - \Omega\left[(2\Omega y - u)\sine{2\Omega t}  + (2\Omega x - v)\cosine{2\Omega t} \right]\p_u  \\
& \qquad + 
\Omega \left[(2\Omega x + v)\sine{2\Omega t} - (2\Omega y + u)\cosine{2\Omega t}\right] \p_v + 2 \Omega h \sine{2\Omega t} \p_h.
\end{split}
\end{equation}
The geometrical interpretation of these generators is as follows.  The system \rref{RSWW-eqs} is left invariant by translations $P_0, P_1, P_2$ in the space of independent variables since it is autonomous.  The element $L$ generates a rotation of the whole coordinate system while $G_1$ and $G_2$ represent helical rotations.  The system is also left invariant by the dilation $D$ and the two conformal transformations $Z_1$ and $Z_2$.

\noindent The Levi decomposition ${\cal L} = F \niplus N$ of the symmetry algebra ${\cal L}$ can be exhibited by considering its commutation table (Table \ref{Commutation_relations}) in the following basis
\begin{equation}
\label{Algebra-basis}
\begin{split}
&Y_1 = P_2 - 2\Omega G_2, \quad Y_2 = -(P_1 + 2\Omega G_1), \quad Y_3 = P_1 - 2\Omega G_1, \quad Y_4 = P_2 + 2\Omega G_2, \quad \\
&Y_5 = -L, \quad Y_6 = D, \quad Y_7 = P_0 - \Omega L - Z_2, \quad Y_8 = P_0 -\Omega L + Z_2, \quad Y_9 = -\frac{1}{\Omega} Z_1. 
\end{split}
\end{equation}
Here $F = \{Y_1,Y_2,Y_3,Y_4,Y_5,Y_6\}$ is a maximal solvable ideal and $N = \{Y_7, Y_8, Y_9\}$ is isomorphic to the simple Lie algebra $su(1,1)$.  Following the procedure presented in \cite{Donato-Oliveri-MHD-1993,Donato-Oliveri-1995}, we introduce a set of canonical variables associated with the abelian subalgebra $\{Y_1,Y_2,Y_7\}$ and defined by
\begin{equation}
\label{Canonical-variables}
\begin{array}{lll}
Y_7 t' = 1,  & Y_1 t' = 0, & Y_2 t' = 0, \\
Y_7 x' = 0,  & Y_1 x' = 1, & Y_2 x' = 0, \\
Y_7 y' = 0,  & Y_1 y' = 0, & Y_2 y' = 1, \\
Y_7 u' = Y_7 v' = Y_7 h' = &Y_1 u' = Y_1 v' = Y_1 h' = &Y_2 u' = Y_2 v' = Y_2 h' = 0,
\end{array}
\end{equation}
to bring system (\ref{RSWW-eqs}) into an equivalent autonomous form.  It turns out that the set of variables (\ref{SWW-RSWW-iso}) satisfies system (\ref{Canonical-variables}) so that when expressed in these variables, the vector fields $Y_1, Y_2, Y_7$ are rectified to the canonical form
$$Y_7 = \p_{t'}, \quad Y_1 = \p_{x'}, \quad Y_2 = \p_{y'}.$$
Moreover, using the chain rule, it is easily found that system (\ref{RSWW-eqs}) transforms to
\begin{equation*}
\begin{split}
&{u'}_{t'} + u' {u'}_{x'} + v' {u'}_{y'} + g {h'}_{x'} = 0,\\
&{v'}_{t'} + u' {v'}_{x'} + v' {v'}_{y'} + g {h'}_{y'} = 0,\\
&{h'}_{t'} + u' {h'}_{x'} + v' {h'}_{y'} + {h'} ({u'}_{x'} + {v'}_{y'}) = 0,
\end{split}
\end{equation*} 
which shows the equivalence between systems (\ref{RSWW-eqs}) and (\ref{SWW-eqs}).  The next section demonstrates how the point transformation (\ref{SWW-RSWW-iso}) can be used to construct implicit solutions of equations (\ref{RSWW-eqs}) expressed in terms of Riemann invariants.

\begin{table}[h]
\begin{equation*}
\begin{array}{c|cccccc|ccc}
&	Y_1	& Y_2	& Y_3	& Y_4	& Y_5	& Y_6	& Y_7		& Y_8	   	&	Y_9 \\\hline
Y_1	& 0	& 0 	& 0	& 0	& -Y_2	& - Y_1 & 0 		& -2\Omega Y_3    	& - Y_1 \\
Y_2	&  	& 0 	& 0	& 0	& Y_1 	& -Y_2 	& 0 		& -2\Omega Y_4     	& - Y_2   \\
Y_3	&  	&      	& 0	& 0 	& -Y_4	& -Y_3 	& 2\Omega Y_1 	& 0		&  Y_3  \\
Y_4	&  	&  	&	& 0	& Y_3 	& -Y_4  & 2\Omega Y_2	& 0     	&  Y_4  \\
Y_5	&  	& 	&	& 	& 0	& 0	& 0        	& 0     	& 0	\\
Y_6	& 	& 	&	& 	& 	& 0	& 0		& 0		& 0	\\\hline
Y_7	& 	& 	&	& 	& 	& 	& 0		& -4\Omega^2 Y_9	& -2 Y_7	\\
Y_8	& 	& 	&	& 	& 	& 	& 		& 0		& 2 Y_8	\\
Y_9	& 	& 	&	& 	& 	& 	& 	 	& 		& 0	\\
\end{array}
\end{equation*}
\caption{Commutation relations for the Lie symmetry algebra of the RSWW equations.}
\label{Commutation_relations}
\end{table}

\section{Conditionally invariant solutions of the SWW and RSWW equations}

We present in this section a brief description of the CSM approach developed progressively in \cite{Grundland07-1} and \cite{Grundland-Huard-JNMP} and obtain several rank-$1$ and rank-$2$ solutions of the SWW equations in closed form.  We illustrate the process of construction of the corresponding solutions for the RSWW equations with several interesting examples.  The SWW equations (\ref{SWW-eqs}) can be written in matrix evolutionary form as
\begin{equation}
\label{SWW-Matrix}
{\bf u}_t + {a}^1({\bf u}) {\bf u}_x + {a}^2({\bf u}) {\bf u}_y = 0,
\end{equation}
where ${a}^1, { a}^2$ are $3 \times 3$ matrix functions given by
$${ a}^1 = \left(\begin{array}{ccc}
	u & 0 & g \\
	0 & u & 0 \\
	h & 0 & u \\
\end{array} \right), \quad { a}^2 = \left(\begin{array}{ccc}
	v & 0 & 0 \\
	0 & v & g \\
	0 & h & v \\
\end{array} \right).$$
The objective is to construct rank-$k$ solutions, $k=1,2$, of system (\ref{SWW-Matrix}) expressible in terms of Riemann invariants.  To this end, we look for solutions of (\ref{SWW-Matrix}) defined implicitly by the relations
\begin{equation}
 \label{rank-k-solution}
\begin{split}
&{\bf u} = {\bf f}(r^1({\bf x},{\bf u}), \ldots, r^k({\bf x},{\bf u})), \quad r^A({\bf x},{\bf u}) = \lambda^A_i({\bf u}) x^i,\\ 
&\det{\left(\lambda_0^A \mathcal{I}_3 + {a}^1({\bf u}) \lambda^A_1 + {a}^2 ({\bf u}) \lambda^A_2\right)} = 0, \quad A=1,\ldots, k,
\end{split}
\end{equation}
for some function $f : \mathbb{R}^k \to \mathbb{R}^3$, where ${\cal I}_3$ is the $3$ by $3$ identity matrix.  A solution of the form (\ref{rank-k-solution}) will be called a rank-$k$ solution if $\mathrm{rank}{(\partial u)} = k$ in some open set $\mathcal{D} \subset \mathbb{R}^3$ around the origin, where $\partial u$ stands for the Jacobian matrix of ${\bf u}$ in the original variables. The functions $r^A(\bfx,\bfu)$ are called the Riemann invariants associated with the linearly independent wave vectors $\lambda^A = (\lambda^A_0, \vec{\lambda}^A) = (\lambda^A_0,\lambda^A_1,\lambda^A_2)$, which are obtained by solving the dispersion relation of equation (\ref{SWW-Matrix}) for the phase velocity $\lambda_0$. This relation takes the form
\begin{equation}
\begin{split}
&\det{\left(\lambda_0 \mathcal{I}_3 + { a}^1({\bf u}) \lambda_1 + {a}^2 ({\bf u}) \lambda_2\right)} \\
&= (\lambda_0 + \lambda_1 u + \lambda_2 v)(\lambda_0 + \lambda_1 u + \lambda_2 v+\sqrt{g h})(\lambda_0 + \lambda_1 u + \lambda_2 v-\sqrt{g h}) = 0.
\end{split}
\end{equation}
The wave vectors are thus of the entropic (E) and acoustic (S) type defined respectively by
\begin{equation}
\label{wave-vectors}
\begin{split}
&\mathrm{i) }\, \lambda^E = (-\lambda_1 u - \lambda_2 v, \lambda_1, \lambda_2),\\
&\mathrm{ii) }\, \lambda^{S_{\varepsilon}} = (-(\lambda_1 u + \lambda_2 v + \varepsilon \sqrt{gh}), \lambda_1, \lambda_2), \quad |\vec{\lambda}|^2 = {\lambda_1}^2+{\lambda_2}^2 = 1, \quad \varepsilon=\pm 1.\\
\end{split}
\end{equation}
We associate to each of them the corresponding Riemann invariant
\begin{equation}
\begin{split}
&\mathrm{i) }\, r^E = -(\lambda_1 u + \lambda_2 v) t + \lambda_1 x + \lambda_2 y,\\
&\mathrm{ii)}\, r^{S_{\varepsilon}} = -(\lambda_1 u + \lambda_2 v + \varepsilon \sqrt{gh}) t + \lambda_1 x + \lambda_2 y, \quad |\vec{\lambda}|^2 = 1, \quad \varepsilon=\pm 1.
\end{split}
\end{equation}
The analysis of rank-$k$ solutions for the cases $\varepsilon=\pm 1$ are very similar, hence we restrict ourselves to the positive case.

It is convenient when studying solutions of type (\ref{rank-k-solution}) to write system (\ref{SWW-Matrix}) in the form of a trace equation,
\begin{equation}
\label{original-system-trace}
\tr{\left[{\cal {A}}^{\mu}({\bf u}) {\partial u}\right]} = 0, \quad \mu = 1,\ldots, l,
\end{equation}
where ${\cal {A}}^{\mu}({\bf u})$ are now $3\times 3$ matrix functions of ${\bf u}$, defined by

$${\cal A}^1 = \left(\begin{array}{ccc}
	1 & 0 & 0 \\
	u & 0 & g \\
	v & 0 & 0 \\
\end{array} \right), \quad 
{\cal A}^2 = \left(\begin{array}{ccc}
	0 & 1 & 0 \\
	0 & u & 0 \\
	0 & v & g \\
\end{array} \right), \quad
{\cal A}^3 = \left(\begin{array}{ccc}
	0 & 0 & 1 \\
	h & 0 & u \\
	0 & h & v \\
\end{array} \right)
.$$
The construction of rank-$k$ solutions through the conditional symmetry method is achieved by considering an overdetermined system, consisting of the original system \rref{SWW-Matrix} together with a set of compatible first order differential constraints (DCs), 
\begin{equation}
\label{Differential-constraints}
\xi_a^i({\bf u}) u^{\alpha}_i = 0, \quad \lambda^A_i({\bf u}) \xi^i_a({\bf u}) = 0, \quad a=1,\ldots,3-k, \quad A = 1, \ldots, k,
\end{equation}
for which a symmetry criterion is automatically satisfied.  Here and throughout this work, we use the summation convention over repeated indices.  Introducing the functions
\begin{equation}
\label{rank-$k$-coord}
\begin{split}
&\bx^1 = r^1(\bfx,\bfu),\, \ldots,\, \bx^k = r^k(\bfx,\bfu), \, \bx^{k+1} = x^{k+1},\, \ldots\\ 
&\bu = u,\, \bar{v} = v,\, \bar{h} = h, \end{split}
\end{equation}
as new coordinates on $\NR^3 \times \NR^3$ space, the Jacobi matrix $\partial u$ now reads 
\begin{equation}
\partial u = \pd{f}{r} \left({\cal I}_k - \left(\eta_0 t + \eta_1 x + \eta_2 y\right) \pd{f}{r}\right)^{-1} \lambda,
\end{equation}
where
\begin{eqnarray}
& & \lambda = (\lambda^A_i) \in
\NR^{k \times 3},\quad r = (r^1,\ldots, r^k) \in
 \NR^k, \quad \frac{\p f}{\p r} = \left(
\frac{\p f^{\alpha}}{\p r^A}\right) \in \NR^{3 \times k}, \nonumber\\
\label{threematrices}
& &\eta_{a} = \left(\pd{\lambda^A_{a}}{u^{\alpha}}\right) \in \NR^{k \times 3}, \quad a=0,\ldots,2,
\end{eqnarray}
so that system (\ref{original-system-trace}) is now expressed as
\begin{equation}
\label{trace-new-coords}
\tr{\left[{\cal {A}}^{\mu}({\bf u}) {\pd{f}{r} \left({\cal I}_k - \left(\eta_0 t + \eta_1 x + \eta_2 y\right) \pd{f}{r}\right)^{-1} \lambda }\right]} = 0, \quad \mu = 1,\ldots, l.
\end{equation}
Requiring that system (\ref{trace-new-coords}) be satisfied for all values of the coordinates $(t,x,y)$, the following result holds (see \cite{Grundland07-1} for a general statement and a detailed proof).
\begin{prop}
The nondegenerate quasilinear hyperbolic system of first order PDEs \rref{SWW-Matrix} admits a $(3-k)$-dimensional conditional symmetry algebra $L$, $k\leq 2$, if and only if there exists a set of $(3-k)$ linearly independent vector fields
\begin{equation*}
X_a = \xi^i_a(u) \pd{}{x^i}, \quad a=1,\ldots,3-k, \quad \det{\left(a^i(u) \lambda^A_i\right)} = 0, \quad \lambda^A_i \xi^i_a = 0, \quad A=1,\ldots, k,
\end{equation*}
which satisfy, on some neighborhood of $(x_0,u_0) \in X \times U$, the trace conditions
\begin{eqnarray}
\label{trace-eq-prop-k-1}
&&\text{$k=1$} :  \quad \mathrm{i)}  \quad \mathrm{tr}{\left({\cal A}^{\mu} \pd{f}{r} \lambda\right)} = 0, \quad  \mu = 1,\ldots, 3,\\
\label{trace-eq-prop-k-2}
&&\text{$k=2$} : \quad  \mathrm{i)}  \quad \mathrm{tr}{\left({\cal A}^{\mu} \pd{f}{r} \lambda\right)} = 0,  \quad
 \mathrm{ii)} \quad \mathrm{tr}{\left({\cal A}^{\mu} \pd{f}{r} \eta_{a} \pd{f}{r} \lambda\right)}=0, \quad a = 0,\ldots, 2,
\end{eqnarray}
where the relevant matrices are defined in (\ref{threematrices}).  Solutions of the system which are invariant under the Lie algebra $L$ are precisely rank-$k$ solutions of the form \rref{rank-k-solution}.
\end{prop}

Note that the vector fields $X_a$, $a=1,\ldots, 3-k$, are not symmetries of the original system.  Nevertheless, as we will show, they can be used to build solutions of the overdetermined system composed of (\ref{SWW-Matrix}) and the differential constraints (\ref{Differential-constraints}).

To construct solutions of the RSWW equations, we assume that a solution of the SWW equations (\ref{SWW-eqs})
$$u = u({\bf r}), \, v= v ({\bf r}), \, h = h({\bf r})\quad {\bf r} = (r^1, \ldots, r^k),$$
has been obtained from equations (\ref{trace-eq-prop-k-1}) or (\ref{trace-eq-prop-k-2}).  Then the Riemann invariants $r^A$ can be expressed as a graph
\begin{equation}
\label{invariants-graph}
r^A = r^A({\bf x},{\bf u}) = r^A({\bf x}, \Phi({\bf r}))
\end{equation}
in the $({\bf r}, {\bf x})$ space for some function $\Phi : \mathbb{R}^k \to \mathbb{R}^q$.  The change of variables (\ref{SWW-RSWW-iso}) induces a transformation of the independent variables in this space,
\begin{equation}
t \to -\frac{1}{2\Omega} \cot{(\Omega t)}, \quad x \to \frac{1}{2} (y - x \cot{(\Omega t)}), \quad y \to -\frac{1}{2} (x+y\cot{(\Omega t)}),
\end{equation}
 and we denote by $\tilde{\bf r} = (\tilde{r}^1, \ldots, \tilde{r}^k)$ the resulting functions in the new variables.  Then, according to transformation (\ref{SWW-RSWW-iso}), the functions
\begin{equation}
\label{SWW-RSWW-transfo}
\begin{split}
&\tilde{u} = - u(\tilde{\bf r}) \cot{\left(\Omega t\right)} - v(\tilde{\bf r}) + \Omega \left(y + x \cot{\left(\Omega\right)}\right),\\
&\tilde{v} = u(\tilde{\bf r}) - v(\tilde{\bf r}) \cot{\left(\Omega t\right)} - \Omega \left(x - y \cot{\left(\Omega t\right)}\right),\\
&\tilde{h} = h(\tilde{\bf r}) \csc^2{\left(\Omega t\right)},
\end{split}
\end{equation}
form a solution of the RSWW equations (\ref{RSWW-eqs}).  Even though tranformation (\ref{SWW-RSWW-iso}) is singular at every time $t = \frac{\pi}{2\Omega}(2n+1)$, $n\in \mathbb{N}$, we show that it is possible to obtain implicit solutions defined in a neigborhood of the origin $t=0$.

\subsection{Rank-1 solutions}

The reduction procedure outlined above has been applied to obtain rank-$1$ and rank-$2$ solutions of the SWW equations (\ref{SWW-eqs}) and their corresponding solutions of the RSWW system (\ref{RSWW-eqs}).  We present here several rank-1 solutions, also called simple waves, associated with the different types of wave vectors (\ref{wave-vectors}). Note that in the case where $k=1$, the CSM and the generalized method of characteristics agree \cite{Grundland07-1}.  

{\bf i) } Simple entropic-type waves are obtained by considering system (\ref{SWW-eqs}) in the new variables
$$\bar{t} = t, \, \bar{x} = r(\bfx,\bfu), \, \bar{y} = y, \quad \bar{u} = u, \bar{v} = v, \bar{h} = h,$$
where $r(\bfx,\bfu) = -(\lambda_1 u + \lambda_2 v) t + \lambda^1_1 x + \lambda^1_2 y$ and the functions $\lambda_i$, $i=1,2$, are allowed to depend on $u,v,h$.  Following Proposition 1, we look for solutions invariant under the vector fields
\begin{equation}
X_1 = \lambda_1 \p_t + (\lambda_1 u + \lambda_2 v) \p_x, \quad X_2 = \lambda_2 \p_t + (\lambda_1 u + \lambda_2 v) \p_y.
\end{equation}
The transformed system (\ref{trace-eq-prop-k-1}) reads as
\begin{equation}
g \lambda_1 h_r = 0, \quad g \lambda_2 h_r = 0, \quad (\lambda_1 u_r + \lambda_2 v_r) h = 0.
\end{equation}
To obtain a nontrivial solution, we must have $h = h_0 \in \mathbb{R}^+$ together with the relation
\begin{equation}
\label{simple-waves-type-1-eq}
\lambda_1 u_r + \lambda_2 v_r = 0.
\end{equation}

For example, if $\lambda_1$ and $\lambda_2$ are constant, we can express $u$ in terms of $v$ and obtain the explicit solution
$$u = u_0 - \frac{\lambda_2}{\lambda_1} v(r), \quad v = v(r), \quad h = h_0,\quad r = - u_0 \lambda_1 t + \lambda_1 x + \lambda_2 y, \quad \lambda_1 \neq 0, \quad h_0 \in \mathbb{R}^{+},$$
where $\lambda_2$ is an arbitrary constant and $v(r)$ is an arbitrary function.  

When the $\lambda_i$ are not constant, different choices can lead to solutions for the velocity vector fields $u(r)$ and $v(r)$ which are of distinct nature.  For example, consider the choice $\lambda_1 = u$, $\lambda_2 = v$, leading to
$$
u u_r + v v_r = \frac{1}{2} (u^2+v^2)_r = 0 \Rightarrow u^2 + v^2 = C^2, \quad C \in \mathbb{R}.
$$
A periodic solution is obtained by choosing
\begin{equation}
\label{Sol-S-periodic}
u = C \sin{r}, \quad v = C \cos{r}, \quad h = h_0, \quad C \in \mathbb{R},
\end{equation}
where the Riemann invariant is given implicitly by
\begin{equation}
\label{Sol-S-RI-periodic}
r = - C( C t - x\sin{r} -  y \cos{r}).
\end{equation}
When $\lambda_1 = v$, $\lambda_2 = u$, equation (\ref{simple-waves-type-1-eq}) implies
$$v u_r + u v_r = (uv)_r = 0 \Rightarrow v = \frac{C}{u(r)},\quad C \in \mathbb{R}.$$
We then get the solution
\begin{equation}
u = u(r), \quad v = \frac{C}{u(r)}, \quad h = h_0 \in\mathbb{R}, \quad r = -2 C t + \frac{C}{u(r)} x + u(r) y,
\end{equation}
where $u(r)$ is an arbitrary function of the Riemann invariant $r$. 

{\bf ii) } Similarly, simple acoustic-type waves are obtained by considering system (\ref{SWW-eqs}) in the new variables
$$\bar{t} = t, \, \bar{x} = r(\bfx,\bfu), \, \bar{y} = y, \quad \bar{u} = u, \bar{v} = v, \bar{h} = h,$$
where $r(\bfx,\bfu) = -(\lambda_1 u + \lambda_2 v + \sqrt{g h}) t + \lambda^1_1 x + \lambda^1_2 y$, $|\vec{\lambda}|^2=1$, and the functions $\lambda_i$, $i=1,2$, are allowed to depend on $u,v,h$.
Rank-1 solutions of this type are invariant under the vector fields
\begin{equation}
X_1 = \lambda_1 \p_t + (\lambda_1 u + \lambda_2 v + \sqrt{g h}) \p_x, \quad X_2 = \lambda_2 \p_t + (\lambda_1 u + \lambda_2 v + \sqrt{g h}) \p_y.
\end{equation}
In this case, the transformed system (\ref{trace-eq-prop-k-1}) is
\begin{equation}
\label{2-reduced-system}
\lambda_1 \sqrt{\frac{g}{h}} h_r = u_r,\quad \lambda_2 \sqrt{\frac{g}{h}} h_r = v_r, \quad h(\lambda_1 u_r + \lambda_2 v_r) = \sqrt{gh} h_r.
\end{equation}
The third equation is automatically satisfied whenever the first two are and $|\vec{\lambda}|^2=1$.  Note that in order to obtain a solution for $h(r)$, it is necessary that the relation
\begin{equation}
\lambda_1(u,v,h) v_r - \lambda_2(u,v,h) u_r = 0
\end{equation}
be satisfied.  Considering different choices for the functions $\lambda_i(u,v,h)$, we obtain several interesting solutions, presented in Table \ref{Table-rank-1-solutions}.

For illustration, we now turn to the construction of the implicit solution of the RSWW equations corresponding to (\ref{Sol-S-periodic}), (\ref{Sol-S-RI-periodic}) using transformation (\ref{SWW-RSWW-iso}).  We first transform the Riemann invariant $r$ to obtain an implicit equation for $\tilde{r}$,
\begin{equation}
\label{simple-wave-type-r-tilde}
\tilde{r} = \frac{C^2}{2\Omega} \cot{\left(\Omega  t\right)} + \frac{C}{2} \left[\left(y - x \cot{\left(\Omega t\right)}\right)\sin{\tilde{r}} - \left(x + y \cot{\left(\Omega t\right)}\right)\cos{\tilde{r}}\right].
\end{equation}
Using equations (\ref{SWW-RSWW-transfo}), we obtain the implicit solution of the RSWW equations
\begin{equation}
\begin{split}
&u = - C \cos{\tilde{r}} - C \cot{\left(\Omega t\right)} \sin{\tilde{r}} + \Omega \left(y + x \cot{\left(\Omega t\right)}\right),\\
&v = C \sin{\tilde{r}} - C \cot{\left(\Omega t\right)}\cos{\tilde{r}}  - \Omega \left(y + x \cot{\left(\Omega t\right)}\right),\\
&h = h_0 \csc^2{\left(\Omega t\right)},
\end{split}
\end{equation}
where $\tilde{r}$ is the solution of the implicit equation (\ref{simple-wave-type-r-tilde}).  This solution has period $\pi / \Omega$ and goes to infinity at every time $t = k \pi / \Omega$, $k \in \mathbb{N}$.  Nevertheless, due to the invariance of equations (\ref{RSWW-eqs}) with respect to translations in time, it is possible to use a time shift $t \to t + t_0$ so that equations (\ref{simple-wave-type-r-tilde}) are well defined in a neighborhood of length $\pi/\Omega$ around $t=0$.  For example, the translation $t \to t + \frac{\pi}{2\Omega}$ gives the solution
\begin{equation}
\begin{split}
&u = - C \cos{\bar{r}} + C \tan{\left(\Omega t\right)} \sin{\bar{r}} + \Omega \left(y - x \tan{\left(\Omega t\right)}\right),\\
&v = C \sin{\bar{r}} + C \tan{\left(\Omega t\right)}\cos{\bar{r}}  - \Omega \left(y - x \tan{\left(\Omega t\right)}\right),\\
&h = h_0 \sec^2{\left(\Omega t\right)},
\end{split}
\end{equation}
where $\bar{r}$ satisfies the equation
\begin{equation}
\bar{r} = -\frac{C^2}{2\Omega} \tan{\left(\Omega  t\right)} + \frac{C}{2} \left[\left(y + x \tan{\left(\Omega t\right)}\right)\sin{\bar{r}} - \left(x - y \tan{\left(\Omega t\right)}\right)\cos{\bar{r}}\right],
\end{equation}
which is clearly defined in the interval $\left(-\frac{\pi}{2\Omega}, \frac{\pi}{2\Omega}\right)$.  Note that this process can be applied to every solution presented in Table \ref{Table-rank-1-solutions} to generate local solutions of the RSWW equations defined around $t=0$.

\subsection{Rank-2 solutions}

The construction of rank-$2$ solutions is much more involved than in the case $k=1$ since it requires us to solve system (\ref{trace-eq-prop-k-2}), which is composed of at most twelve independent nonlinear partial differential equations, compared to only three equations.  However, we now show that the task is undertakable and leads to interesting solutions. The results of this analysis are summarized in Table \ref{Table-Rank-2-SWW} and \ref{Table-Rank-2-RSWW}. 

{\bf i) } We first look for rank-$2$ solutions resulting from the interaction of two entropic-type solutions. They are invariant under the vector field
\begin{equation}
X = \p_t + u \p_x + v \p_y.
\end{equation}
In the variables\\
\begin{equation}
\begin{split}
&\bar{t} = t, \, \bar{x}^1 = r^1(\bfx,\bfu), \, \bar{x}^2 = r^2(\bfx,\bfu), \, \bar{u}=u, \, \bar{v} = v,\, \bar{h} = h,\\
&r^i(\bfx,\bfu) = t - \frac{\lambda^i_1}{\lambda^i_1 u + \lambda^i_2 v} x - \frac{\lambda^i_2}{\lambda^i_1 u + \lambda^i_2 v} y, \quad i = 1,2,
\end{split}
\end{equation}
equations (\ref{trace-eq-prop-k-2} i) read as
\begin{eqnarray}
&&\lambda^1_1 (\lambda^2_1 u + \lambda^2_2 v) h_{r^1} + \lambda^2_1 (\lambda^1_1 u + \lambda^1_2 v) h_{r^2} = 0,\\
&&\lambda^1_2 (\lambda^2_1 u + \lambda^2_2 v) h_{r^1} + \lambda^2_2 (\lambda^1_1 u + \lambda^1_2 v) h_{r^2} = 0,\\
\label{1-1-third-equation}
&&(\lambda^2_1 u + \lambda^2_2 v) (\lambda^1_1 u_{r^1} + \lambda^1_2 v_{r^1}) + (\lambda^1_1 u + \lambda^1_2 v) (\lambda^2_1 u_{r^2} + \lambda^2_2 v_{r^2}) = 0.
\end{eqnarray}
A solution to the first two equations exists if and only if
$$(\lambda^1_1 \lambda^2_2 - \lambda^1_2 \lambda^2_1)(\lambda^1_1 u + \lambda^1_2 v)(\lambda^2_1 u + \lambda^2_2 v) = 0 \quad \text{or} \quad h = h_0 \in \mathbb{R}^+.$$
The conditions on the functions $\lambda^i_j$ imply either that the wave vectors are parallel or one of the considered waves has zero velocity.  From these conditions, we now show that no rank-$2$ solution can be built from this type of interaction.

When $\vec{\lambda}^2 = k \vec{\lambda}^1$, the Riemann invariants $r^1$ and $r^2$ are equal, hence the solution cannot be of rank $2$.
Therefore we look for solutions with $h = h_0$, a positive constant. 
Equation (\ref{1-1-third-equation}) implies that
\begin{equation}
\label{1-1-hyp-ur1}
u_{r^1} = -\frac{1}{\lambda^1_1} \frac{\lambda^1_1 u + \lambda^1_2 v}{\lambda^2_1 u + \lambda^2_2 v} \left(\lambda^2_1 u_{r^2} + \lambda^2_2 v_{r^2} \right) - \frac{\lambda^1_2}{\lambda^1_1} v_{r^1}.
\end{equation}
We then consider the linear combination
\begin{equation}
\begin{split}
&\frac{1}{uv} \tr{\left[{\cal A}^3 \frac{\p f}{\p r} \left( u \eta_1 + v \eta_2\right)\frac{\p f}{\p r} \lambda\right]} =
 -\frac{2}{uv} \left(\lambda^1_1 u + \lambda^1_2 v\right)\left(\lambda^2_1 u + \lambda^2_2 v\right)\left(\lambda^1_1 \lambda^2_2 - \lambda^1_2 \lambda^2_1\right) \times \\
&\quad \left(\left(\lambda^1_1 u + \lambda^1_2 v\right)\left(\lambda^2_1 u_{r^2} + \lambda^2_2 v_{r^2}\right) v_{r^2} + \left(\lambda^2_1 u + \lambda^2_2 v\right) \left( \lambda^1_1 u_{r^2} + \lambda^1_2 v_{r^2}\right) v_{r^1}\right),
\end{split}
\end{equation}
implying that a rank-2 solution must satisfy 
\begin{equation}
\label{1-1-linear-combination}
\left(\lambda^1_1 u + \lambda^1_2 v\right)\left(\lambda^2_1 u_{r^2} + \lambda^2_2 v_{r^2}\right) v_{r^2} + \left(\lambda^2_1 u + \lambda^2_2 v\right) \left( \lambda^1_1 u_{r^2} + \lambda^1_2 v_{r^2}\right) v_{r^1} = 0.
\end{equation}
When $\lambda^2_1 u_{r^2} + \lambda^2_2 v_{r^2} = 0$, equation (\ref{1-1-hyp-ur1}) requires that
$$u_{r^1} = -\frac{\lambda^1_2}{\lambda^1_1} v_{r^1}, \quad v_{r^2} = -\frac{\lambda^2_1}{\lambda^2_2} u_{r^2},$$
so that (\ref{1-1-linear-combination}) becomes
$$\frac{1}{\lambda^2_2} (\lambda^2_1 u + \lambda^2_2 v)(\lambda^1_1 \lambda^2_2 - \lambda^2_1 \lambda^1_2) u_{r^2} v_{r^1} = 0,$$
leading necessarily to a rank-$1$ solution.  Hence we can solve (\ref{1-1-linear-combination}) for $v_{r^2}$, and the expression (\ref{1-1-hyp-ur1}) for $u_{r^1}$  implies that
\begin{equation}
\frac{u_{r^2}}{u_{r^1}} = \frac{v_{r^2}}{v_{r^1}} = -\frac{\left(\lambda^2_1 u + \lambda^2_2 v\right)\left(\lambda^1_1 u_{r^2} + \lambda^1_2 v_{r^2}\right)}{\left(\lambda^1_1 u + \lambda^1_2 v\right)\left(\lambda^2_1 u_{r^2} + \lambda^2_2 v_{r^2}\right)},
\end{equation}
hence we must have $v = F(u)$, for an arbitrary function $F : \mathbb{R} \to \mathbb{R}$.  But this implies that the Jacobian matrix of the solution is of rank 1, since $h=h_0$.  Thus, no rank-2 solution of type E-E exists.  For example, consider the simplest case when $\lambda^1_1 = \lambda^2_2 = 1, \lambda^1_2 = \lambda^2_1 = 0$.  The Riemann invariants are then given by
$$r^1 = t - \frac{x}{u}, \quad r^2 = t - \frac{y}{v}.$$
Equations (\ref{1-1-hyp-ur1}) and (\ref{1-1-linear-combination}) become
\begin{equation}
\label{1-1-example-reduced}
u v_{r^2} + v u_{r^1} = 0, \quad u {v_{r^2}}^2 + v u_{r^2} v_{r^1} = 0.
\end{equation}
Solving for $v_{r^1}$ and $v_{r^2}$, we obtain that the rank of the Jacobian matrix 
\begin{equation}
J = \frac{\p (u,v,h)}{\p (r^1,r^2)} = \left(\begin{array}{cc}u_{r^1} & u_{r^2} \\ -\frac{v}{u} \frac{{u_{r^1}}^2}{u_{r^2}} & -\frac{v}{u} u_{r^1} \\ 0 & 0 \end{array}\right)
\end{equation}
is equal to one.  A particular solution of (\ref{1-1-example-reduced}) is given by
\begin{equation} 
\begin{split}
&u = (-1)^m s^m, \quad m \neq -1,\quad s = \frac{C_1 r^2 + C_2}{C_3 r^1 + C_4}, \quad C_i \in \mathbb{R},\quad i=1,\ldots,5,\\
&v = C_5 \mathrm{exp} \left(\frac{C_3}{C_1} m s \right), \quad h = h_0 \in \mathbb{R}^{+},
\end{split}
\end{equation}
which is indeed seen to depend on the single variable $s$.

{\bf ii)}  We now look for interactions of a solution of each type.  This type of solution is invariant under the vector field
\begin{equation}
X = \delta \p_t + \left( \delta u - \lambda^1_2 \sqrt{g h}\right) \p_x  + \left( \delta v - \lambda^1_1 \sqrt{g h}\right) \p_y, \quad \delta = \lambda^1_1 \lambda^2_2 - \lambda^1_2 \lambda^2_1.
\end{equation}
Introducing the change of variables
$$\bar{t} = t, \, \bar{x}^1 = r^1(\bfx,\bfu), \, \bar{x}^2 = r^2(\bfx,\bfu), \, \bar{u}=u, \, \bar{v} = v,\, \bar{h} = h,$$
with
\begin{equation}
\begin{split}
&r^1(\bfx,\bfu) = t - \frac{\lambda^1_1}{\lambda^1_1 u + \lambda^1_2 v} x - \frac{\lambda^2_1}{\lambda^1_1 u + \lambda^1_2 v} y,\\
&r^2(\bfx,\bfu) = t - \frac{\lambda^2_1}{\lambda^2_1 u + \lambda^2_2 v + \sqrt{g h}} x - \frac{\lambda^2_2}{\lambda^2_1 u + \lambda^2_2 v + \sqrt{g h}} y,
\end{split}
\end{equation}
we show that rank-2 solutions can be built by setting $\lambda^2_1 = 1, \lambda^2_2=0$.  Supposing that $\lambda^1_1,\lambda^1_2 \neq 0$, equations (\ref{trace-eq-prop-k-2}) require that
\begin{equation}
\label{1-2-a-reduced}
\begin{split}
&u_{r^1} = -\frac{{\lambda^1_1}^2 - {\lambda^1_2}^2}{\lambda^1_2 ({\lambda^1_1}^2 + {\lambda^1_2}^2)} v_{r^2}, \quad v_{r^1} = - \frac{2 \lambda^1_1}{{\lambda^1_1}^2 + {\lambda^1_2}^2} v_{r^2}, \\
&h_{r^1} = \frac{\sqrt{g h}}{\lambda^1_2 g} v_{r^2},  \quad h_{r^2} = \frac{\sqrt{g h}}{\lambda^1_2 g} (\lambda^1_2 u_{r^2} - \lambda^1_1 v_{r^2}), \\
&(3 {\lambda^1_2}^2 - {\lambda^1_1}^2)( ({\lambda^1_2}^2 - {\lambda^1_1}^2) v_{r^2} + 2\lambda^1_1 \lambda^1_2 u_{r^2}) = 0,\\
&( ({\lambda^1_2}^2 - {\lambda^1_1}^2) v_{r^2} + 2\lambda^1_1 \lambda^1_2 u_{r^2})(\sqrt{g h} {\lambda^1_1}_{,u} + h {\lambda^1_1}_{,h}) = 0, \\ 
&(({\lambda^1_2}^2 - {\lambda^1_1}^2) v_{r^2} + 2\lambda^1_1 \lambda^1_2 u_{r^2})(\sqrt{g h} {\lambda^1_2}_{,u} + h {\lambda^1_2}_{,h}) = 0.
\end{split}
\end{equation}
It is easily computed from equations (\ref{1-2-a-reduced}) that when $({\lambda^1_2}^2 - {\lambda^1_1}^2) v_{r^2} + 2\lambda^1_1 \lambda^1_2 u_{r^2} = 0,$ the obtained solution will be of rank 1.  Thus, we must have $\lambda^1_1 = F_1(u - 2 \sqrt{g h}, v)$, $\lambda^1_2 = F_2(u - 2\sqrt{g h}, v)$ where $F_1, F_2$ are arbitrary functions.
Equations (\ref{1-2-a-reduced}) can be solved for specific choices of the arbitrary functions $F_1, F_2$.  Hence we consider the case where $({\lambda^1_2}^2 - {\lambda^1_1}^2) v_{r^2} + 2\lambda^1_1 \lambda^1_2 u_{r^2} \neq 0$, 
together with the relation  $\lambda^1_1 = \sqrt{3}\varepsilon {\lambda^1_2}$, $\varepsilon = \pm 1$, which leads to
\begin{eqnarray}
\label{1-2-a-1}
&&u_{r^1} = -\frac{1}{2 \lambda^1_2} v_{r^2}, \\
\label{1-2-a-equation-v}
&&v_{r^1} = -\frac{\sqrt{3} \varepsilon}{2 \lambda^1_2} v_{r^2}, \\
\label{1-2-a-3}
&&h_{r^1} = \frac{1}{\lambda^1_2}\sqrt{\frac{h}{g}}  v_{r^2},\quad  h_{r^2} = \sqrt{\frac{h}{g}} (u_{r^2} - \sqrt{3} \varepsilon v_{r^2}),\\
\label{1-2-a-4}
&&\sqrt{g h} {\lambda^1_2}_{,u} + h {\lambda^1_2}_{,h} = 0.
\end{eqnarray}
When $\lambda^1_2$ is a function of $v$ only, system (\ref{1-2-a-1}) - (\ref{1-2-a-4}) is compatible and can be integrated to yield
\begin{equation}
u = \frac{\sqrt{3}}{3} \varepsilon v(r^1,r^2) + F(r^2), \quad h = \frac{1}{4 g} \left( F(r^2) - \frac{2\sqrt{3}}{3} \varepsilon v(r^1,r^2) + h_0 \right)^2,
\end{equation}
where $v(r^1,r^2)$ is given implicitly by
\begin{equation}
v = G\left(s\right), \quad s = r^2 - \frac{\sqrt{3} \varepsilon}{2 \lambda^1_2 (v)} r^1,
\end{equation}
and $G(s)$ is an arbitrary function of its argument.  The Riemann invariants $r^1$ and $r^2$ then satisfy the implicit relations
\begin{equation}
\label{1-2-RIs}
\begin{split}
&r^1 = \lambda^1_2(G(s)) \left( \left(2 G(s) + \sqrt{3} \varepsilon F(r^2)\right)t - \sqrt{3} \varepsilon x - y\right), \quad s = r^2 - \frac{\sqrt{3} \varepsilon}{2 \lambda^1_2 (v)} r^1,\quad \varepsilon = \pm 1,\\
&r^2 = \left(\frac{3}{2} F(r^2) + \frac{h_0}{2}\right) t - x.
\end{split}
\end{equation}
Because of the nonlinear coupling of the Riemann invariants (\ref{1-2-RIs}), this type of solution is said to be scattering.  For different choices of the function $\lambda^1_2$ and the profile of $v$ (i.e. $G(s)$), following the construction presented in \cite{Conte-Grundland-Huard-JPhysA-2009}, it is possible to construct rank-2 solutions which are bounded everywhere, for example bumps, kinks and periodic solutions, even when the Riemann invariants admit the gradient catastrophe after a finite time. We present in Table \ref{Table-Rank-2-Bounded} several solutions of the SWW equations obtained in this way. According to equations (\ref{SWW-RSWW-iso}), after a time shift $t \to t + \pi/2\Omega$, the RSWW equations admit the following solution
\begin{equation}
\label{SE-RSWW-Sol}
\begin{split}
u &= - \left( \frac{\sqrt{3}}{3} \varepsilon G(\tilde{s}) + F(\tilde{r}^2) \right) \tan{\left(\Omega t\right)} - G(\tilde{s}) + \Omega (y+x \tan{(\Omega t)}), \quad \varepsilon = \pm 1,\\
v &= \frac{\sqrt{3}}{3} \varepsilon G(\tilde{s}) + F(\tilde{r}^2) - G(\tilde{s}) \tan{(\Omega t)} - \Omega (x - y \tan{(\Omega t)}),\\
h &= \frac{1}{4 g} \left( F(\tilde{r}^2) - \frac{2 \sqrt{3}}{3} \varepsilon G(\tilde{s}) + h_0 \right)^2 \sec^2{(\Omega t)}, 
\end{split}
\end{equation}
where the functions $\tilde{r}^1, \tilde{r}^2, \tilde{s}$ now satisfy the implicit relations
\begin{equation}
\label{SE-RSWW-RI}
\begin{split}
\tilde{r}^1 &= \lambda^1_2\left(G(\tilde{s})\right) \left( -\frac{1}{2\Omega} (2 G(\tilde{s}) + \sqrt{3} \varepsilon F(\tilde{r}^2)) \tan{(\Omega t)} - \frac{\sqrt{3}}{2} \varepsilon (y - x \tan{(\Omega t)}) + \frac{1}{2} (x + y \tan{(\Omega t)}) \right),\\
\tilde{r}^2 &= - \frac{1}{4\Omega} \left( 3 F(\tilde{r}^2) + h_0 \right) \tan{(\Omega t)} - \frac{1}{2} (y - x \tan{(\Omega t)}), \quad \tilde{s} = \tilde{r}^2 - \frac{\sqrt{3} \varepsilon}{2 \lambda^1_2 \left(G(\tilde{s})\right)} \tilde{r}^1, 
\end{split}
\end{equation}
and $G(\tilde{s})$, $\lambda^1_2(G(\tilde{s}))$ and $F(\tilde{r}^2)$  are arbitrary functions of their respective argument.  Equations (\ref{SE-RSWW-Sol}) and (\ref{SE-RSWW-RI}) define a rank-$2$ solution in the interval $\left(-\frac{\pi}{2\Omega}, \frac{\pi}{2\Omega}\right)$. From bounded solutions of the SWW equations (see Table \ref{Table-Rank-2-Bounded}), one can then construct rank-$2$ solutions of the RSWW equations which are bounded in this interval.

{\bf iii)  } We now turn to the analysis of the interaction of two acoustic-type solutions. Therefore, introducing the change of variables
$$\bar{t} = t, \, \bar{x}^1 = r^1(\bfx,\bfu), \, \bar{x}^2 = r^2(\bfx,\bfu), \, \bar{u}=u, \, \bar{v} = v,\, \bar{h} = h,$$
with
$$r^i(\bfx,\bfu) = -(\lambda^i_1 u + \lambda^i_2 v + \sqrt{g h}) t + \lambda^i_1 x + \lambda^i_2 y, \quad |\vec{\lambda}^i|^2 = 1,\quad i=1,2,$$
the system (\ref{trace-eq-prop-k-2}) is formed of twelve independent equations.  Equations (\ref{trace-eq-prop-k-2} i) are in this case
\begin{eqnarray}
\label{EE-a}
&&g(\lambda^1_1 h_{r^1} + \lambda^2_1 h_{r^2}) = \sqrt{g h}(u_{r^1} + u_{r^2}),\\
\label{EE-b}
&&g(\lambda^1_2 h_{r^1} + \lambda^2_2 h_{r^2}) = \sqrt{g h}(v_{r^1} + v_{r^2}),\\
\label{EE-c}
&&h(\lambda^1_1 u_{r^1} + \lambda^2_1 u_{r^2} + \lambda^1_2 v_{r^1} + \lambda^2_2 v_{r^2}) = \sqrt{g h} (h_{r^1} + h_{r^2}).
\end{eqnarray}
A process of elimination of the derivatives of the functions $\lambda^i_j(u,v,h)$ in (\ref{trace-eq-prop-k-2}ii), leads us to a system composed of
\begin{equation}
\label{2-2-last-eq}
\begin{split}
(u_{r^1} + u_{r^2})(\lambda^1_2 u_{r^1} - \lambda^1_1 v_{r^1} + \lambda^2_2 u_{r^2} - \lambda^2_1 v_{r^2}) = 0,\\
(v_{r^1} + v_{r^2})(\lambda^1_2 u_{r^1} - \lambda^1_1 v_{r^1} + \lambda^2_2 u_{r^2} - \lambda^2_1 v_{r^2}) = 0, 
\end{split}
\end{equation}
and a third complicated expression which takes a much simpler form depending on the branch of solution chosen in (\ref{2-2-last-eq}). \\

{\it a)} If $u(r^1, r^2) = F(r^1-r^2)$ and $v(r^1,r^2) = G(r^1-r^2)$, then the last equation is automatically satisfied. The solution is obtained from the system
\begin{equation}
(\lambda_1^1 - \lambda^2_1) F' + (\lambda^1_2 - \lambda^2_2) G' = 0, \quad |\vec{\lambda}^i|^2 = 1, \quad h = h_0 \in \mathbb{R}^+.
\end{equation}
However, it should be noted that any solution built from this branch reduces to a rank-$1$ entropic-type solution.  Indeed, since $h = h_0$, by equations (\ref{EE-a}) and (\ref{EE-b}), the Jacobian matrix of the solution in the original variables reads as
$$\left| 
\begin{array}{cc} 
F'(r^1-r^2) (r^1_t-r^2_t) & -F'(r^1-r^2) (r^1_x-r^2_x) \\
G'(r^1-r^2) (r^1_t-r^2_t) & -G'(r^1-r^2) (r^1_x-r^2_x)\\
0 & 0
\end{array}
\right|,
$$
which is manifestly of rank-$1$.  Moreover, it can be easily seen that the resulting rank-$1$ solution will be a solution of the first type.  For example, choosing
$$\lambda^1_1 = 1, \quad \lambda^1_2 = 0, \quad \lambda^2_1 = \frac{1-u^2}{1+u^2}, \quad \lambda^2_2 = \frac{2 u}{u^2+1},$$
we obtain the solution
$$v = v_0 + \frac{1}{2} u^2, \quad h = h_0,$$
where $u = F(s)$ is an arbitrary function of 
$$s = r^1 - r^2 = - \frac{F(F^2-2 v_0)}{1+F^2} t + \frac{2 F^2}{1+F^2} x - \frac{2 F}{1+F^2} y.$$

\begin{figure}[h]
\centering \includegraphics[width=8cm]{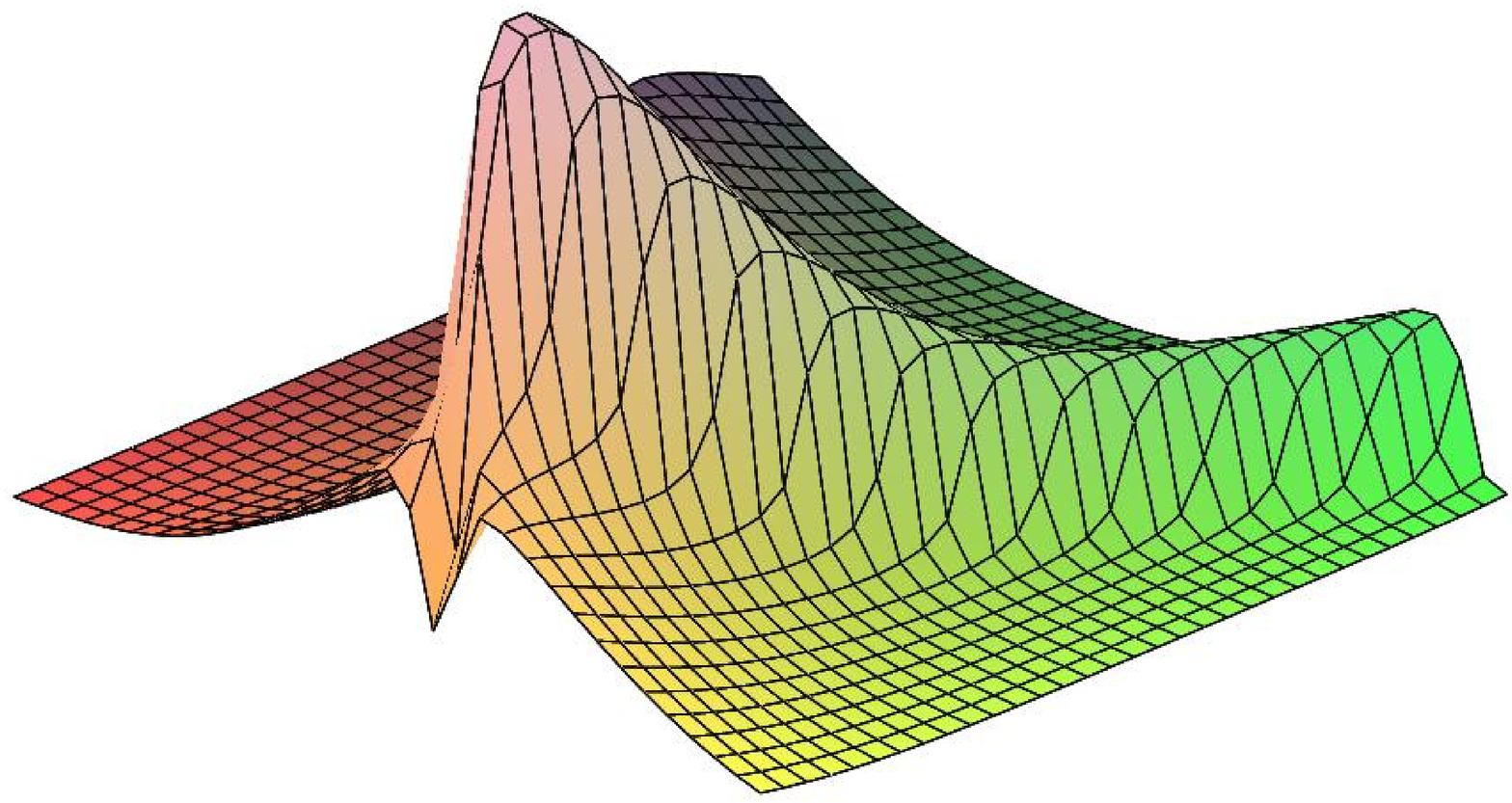} \includegraphics[width=8cm]{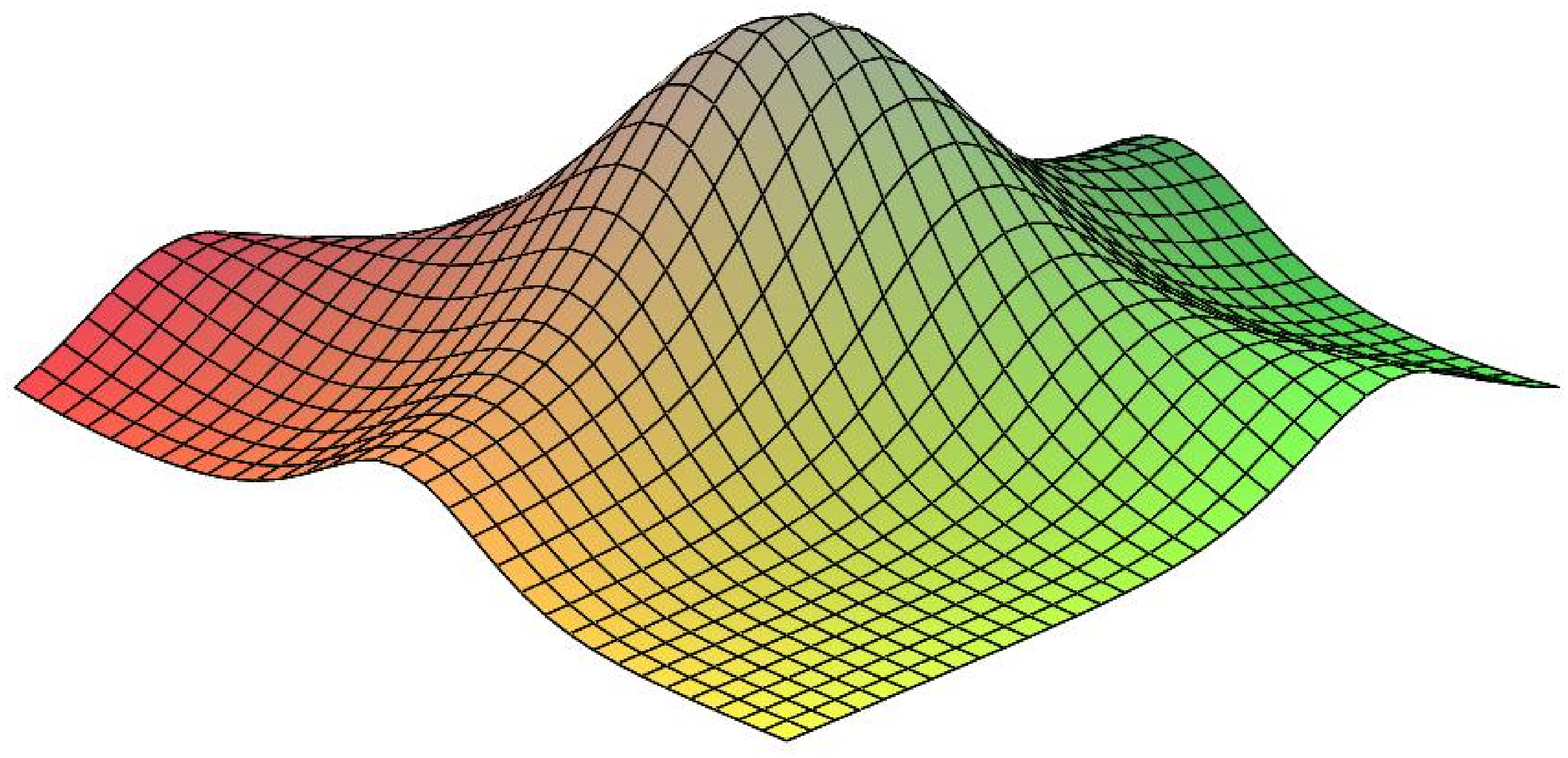}
\caption{Graph of the height function $h(t,x,y)$ for the rank-$2$ solution of the SS type (\ref{Height-Function}) at times $t=-\pi/5$ and $t=0$.}
\label{EE-height-fig}
\end{figure}

{\it b)} When $\lambda^1_2 u_{r^1} - \lambda^1_1 v_{r^1} + \lambda^2_2 u_{r^2} - \lambda^2_1 v_{r^2} = 0$, the last equation reduces to
\begin{equation}
\label{2-2-b-last-equation}
\left[2 \delta^2 + (\vec{\lambda}^1 \cdot \vec{\lambda}^2) - 1\right] v_{r^1} v_{r^2} = 0, \quad \delta = \left|\begin{array}{cc} \lambda^1_1 & \lambda^1_2 \\ \lambda^2_1 & \lambda^2_2 \end{array}\right|, \quad |\vec{\lambda}^i|^2 = 1.
\end{equation}
The solution is necessarily of rank-$1$ if $v_{r^1}=0$ or $v_{r^2}=0$.  We then suppose that $v$ depends essentially on $r^1$ and $r^2$, so the wave vectors $\vec{\lambda}^1$ and $\vec{\lambda}^2$ must satisfy the relations
\begin{equation}
\label{EE-relation-wave-vectors}
 2 \delta^2 + (\vec{\lambda}^1 \cdot \vec{\lambda}^2) - 1 = 0, \quad |\vec{\lambda}^i|^2 = 1, \quad i=1,2.
\end{equation}
Writing $\lambda^1_1 = \sin{\varphi_1}, \lambda^1_2 = \cos{\varphi_1}, \lambda^2_1 = \sin{\varphi_2}, \lambda^2_2 = \cos{\varphi_2}$, equation (\ref{EE-relation-wave-vectors}) implies that the angle $\varphi = |\varphi_1 - \varphi_2| \in [0,2\pi)$ between the wave vectors $\vec{\lambda}^1$ and $\vec{\lambda}^2$ has to satisfy
$$2 \sin^2{\varphi} + \cos{\varphi} - 1 = 0, $$
which can be written as
$$-2 \left(\cos{\varphi} + \frac{1}{2} \right) \left(\cos{\varphi} - 1 \right) =0.$$
Therefore, excluding the case where $\varphi = 0$, we obtain
\begin{equation}
\label{E-E-angle-var}
\cos{\varphi} = \vec{\lambda}^1 \cdot \vec{\lambda}^2 = -1/2 \quad \Rightarrow \quad \varphi = |\varphi_1 - \varphi_2| = 2 \pi/3,
\end{equation}
in accordance with results already obtained for an isentropic fluid flow \cite{Grundland07-1, Peradzynski-GMC}.  In this case, since by (\ref{EE-relation-wave-vectors}) and (\ref{E-E-angle-var}) we must have $\delta = \varepsilon \sqrt{3}/2$, $\varepsilon = \pm 1$,  one can show that the system composed of (\ref{EE-a}) - (\ref{2-2-last-eq}) becomes
\begin{equation}
\label{EE-solved}
\begin{split}
u_{r^1} = \lambda^1_1 \sqrt{\frac{g}{h}} h_{r^1} , \quad u_{r^2} = \lambda^2_1 \sqrt{\frac{g}{h}} h_{r^2}, \quad v_{r^1} = \lambda^1_2 \sqrt{\frac{g}{h}} h_{r^1}, \quad v_{r^2} = \lambda^2_2 \sqrt{\frac{g}{h}} h_{r^2},
\end{split}
\end{equation}
and that the functions $\lambda^i_j$ must satisfy the equations
\begin{equation}
\label{2-2-traces-eqs-c}
\lambda^1_1 {\lambda^2_1}_{,u} + \lambda^1_2 {\lambda^2_1}_{,v} + \frac{h}{\sqrt{g h}} {\lambda^2_1}_{,h} = 0, \quad \lambda^2_1 {\lambda^1_1}_{,u} + \lambda^2_2 {\lambda^1_1}_{,v} + \frac{h}{\sqrt{g h}} {\lambda^1_1}_{,h} = 0.
\end{equation}
Using (\ref{2-2-traces-eqs-c}) and writing $h = H(r^1,r^2)^2$, the compatibility conditions of equations (\ref{EE-solved}) yield the relation
\begin{eqnarray}
H_{r^1 r^2} = 0  \Rightarrow  h(r^1, r^2) = \left(h_1(r^1) + h_2(r^2)\right)^2.
\end{eqnarray}
When the velocity vectors $\vec{\lambda}^1$ and $\vec{\lambda}^2$ are constant, integration of (\ref{EE-solved}) then shows that the velocity vector fields split as a linear sum.  Hence, we obtain the nonscattering solution
\begin{equation}
\label{EE-sol-nonscattering}
\begin{split}
&u = u_0 + 2\sqrt{g} \left(\lambda^1_1 h_1(r^1) + \lambda^2_1 h_2(r^2)\right), \quad v = v_0 + 2\sqrt{g} \left(\lambda^1_2 h_1(r^1) + \lambda^2_2 h_2(r^2)\right), \\
&h = \left(h_1(r^1) + h_2(r^2)\right)^2,
\end{split}
\end{equation}
where the functions $h_1(r^1)$ and $h_2(r^2)$ are arbitrary functions of the Riemann invariants
\begin{equation}
\label{EE-RI-nonscattering}
\begin{split}
&r^1 = - \left(\lambda^1_1 u_0 + \lambda^1_2 v_0 + 3 \sqrt{g} h_1(r^1)\right) t + \lambda^1_1 x + \lambda^1_2 y, \quad \lambda^i_j \in \mathbb{R}, \quad |\vec{\lambda}^i|^2 = 1,\\  
&r^2 = - \left(\lambda^2_1 u_0 + \lambda^2_2 v_0 + 3 \sqrt{g} h_2(r^2)\right) t + \lambda^2_1 x + \lambda^2_2 y, \quad \vec{\lambda}^1 \cdot \vec{\lambda}^2 = -1/2,
\end{split}
\end{equation}
so that the angle between the vectors $\vec{\lambda}^1$ and $\vec{\lambda}^2$ is fixed by relation (\ref{E-E-angle-var}).  Once more, these arbitrary functions can be selected as to ensure that the solution remains bounded everywhere, see Table \ref{Table-Rank-2-Bounded}. By means of transformation (\ref{SWW-RSWW-iso}), we obtain the solution of the RSWW equations (\ref{RSWW-eqs}) corresponding to solution (\ref{EE-sol-nonscattering}).  It is given by 
\begin{equation}
\label{EE-transformed-sol}
\begin{split}
u &= -(u_0 +2 \sqrt{g}(\lambda^1_1 h_1(\tilde{r}^1) + \lambda^2_1 h_2(\tilde{r}^2)) \cot{(\Omega t)} - (v_0 + 2 \sqrt{g} (\lambda^1_2 h_1(\tilde{r}^1) + \lambda^2_2 h_2(\tilde{r}^2))) + \Omega (y+x \cot{(\Omega t)}),\\
v &= u_0 +2 \sqrt{g}(\lambda^1_1 h_1(\tilde{r}^1) + \lambda^2_1 h_2(\tilde{r}^2)) - (v_0 + 2 \sqrt{g} (\lambda^1_2 h_1(\tilde{r}^1) + \lambda^2_2 h_2(\tilde{r}^2))) \cot{(\Omega t)} - \Omega (x - y \cot{(\Omega t)}),\\
h &= (h_1(\tilde{r}^1) + h_2(\tilde{r}^2)) \csc^2({\Omega t}),
\end{split} 
\end{equation}
where the transformed Riemann invariants $\tilde{r}^1, \tilde{r}^2$ satisfy the implicit relations
\begin{equation}
\label{EE-transformed-RI}
\begin{split}
\tilde{r}^1 = \frac{1}{2} \left[ \frac{1}{\Omega}(\lambda^1_1 u_0 + \lambda^1_2 v_0 + 3 \sqrt{g} h_1(\tilde{r}^1)) \cot{(\Omega t)} + \lambda^1_1 (y- x \cot{(\Omega t)}) - \lambda^1_2 (x + y \cot{(\Omega t)})\right],\\
\tilde{r}^2 = \frac{1}{2} \left[ \frac{1}{\Omega}(\lambda^2_1 u_0 + \lambda^2_2 v_0 + 3 \sqrt{g} h_2(\tilde{r}^2)) \cot{(\Omega t)} + \lambda^2_1 (y- x \cot{(\Omega t)}) - \lambda^2_2 (x + y \cot{(\Omega t)})\right].
\end{split}
\end{equation}
Again, it is interesting to note that due to the invariance of equations (\ref{RSWW-eqs}) with respect to translations in time, it is possible to use a time translation $t \to t + t_0$ so that equations (\ref{EE-transformed-RI}) are well defined for $t=0$. For example, when functions $h_1(r^1)$, $h_2(r^2)$ are assumed to be hyperbolic functions of their respective argument, i.e. $h_1(r^1) = \sech^2{(r^1)}$, $h_2(r^2) = \sech^2{(r^2)}$, and if we choose $\vec{\lambda}^1 = (1,0)$ and $\vec{\lambda^2} = (-1/2, \sqrt{3}/2)$, then we obtain after a time shift $t \to t + \pi/2\Omega$ the singular bump-type solution
\begin{eqnarray}
&&\hspace{-2cm}
u = \left(u_0 + \sqrt{g} (2 \sech^2(r1) - \sech^2(r^2))\right) \tan{(\Omega t)} - (v_0 + \sqrt{3 g} \sech^2(r^2)) + \Omega(y - x \tan{(\Omega t)}) , \nonumber\\
&&\hspace{-2cm}
v = (v_0 + \sqrt{3 g} \sech^2(r^2)) \tan{(\Omega t)} + \sqrt{g} (2 \sech^2(r^1) - \sech^2(r^2)) - \Omega (x + y \tan{(\Omega t)}), \nonumber\\
&&\hspace{-2cm}
\label{Height-Function}
h = (\sech^2(r^1) + \sech^2(r^2))^2 \sec^2(\Omega t)
\end{eqnarray}
with
\begin{eqnarray}
\label{E-E-graph-RI}
&&\hspace{-2cm}
r^1 = \frac{1}{2\Omega} \left(u_0 + 3 \sqrt{g} \sech^2(r^1) \right) \tan{(\Omega t)} + \frac{1}{2} \left(y - x \tan{(\Omega t)}\right), \\
&&\hspace{-2cm}
r^2 = \frac{1}{2\Omega} \left(\frac{u_0}{2} + \frac{\sqrt{3}}{2} v_0 + 3 \sqrt{g} \sech^2(r^2) \right) \tan{(\Omega t)} - \frac{1}{4} \left( y - x \tan{(\Omega t)} \right) - \frac{\sqrt{3}}{4} \left( x + y \tan{(\Omega t)} \right)\nonumber.
\end{eqnarray}
Figure \ref{EE-height-fig} illustrates the behavior of the height function $h(t,x,y)$ defined by (\ref{Height-Function}) and (\ref{E-E-graph-RI}) .   

When the $\lambda^i_j$ are not constant,  equations (\ref{2-2-traces-eqs-c}) possess several classes of implicit solutions.  Supposing that 
\begin{equation}
\label{EE-nc-1112}
\lambda^1_1 = \frac{\Psi}{\sqrt{1+\Psi^2}}, \quad \lambda^1_2 = \frac{1}{\sqrt{1+\Psi^2}},
\end{equation}
for some function $\Psi : \mathbb{R}^3 \to \mathbb{R}$, equation (\ref{2-2-b-last-equation}) requires that
\begin{equation}
\label{EE-nc-2122}
\lambda^2_1 = -\frac{1}{2} \frac{\Psi+\sqrt{3}}{\sqrt{1+\Psi^2}}, \quad \lambda^2_2 = \frac{1}{2} \frac{\sqrt{3} \Psi - 1}{\sqrt{1+\Psi^2}}.
\end{equation}
The system (\ref{2-2-traces-eqs-c}) then becomes
\begin{equation}
\label{2-2-system-for-Psi}
\begin{split}
&\Psi \Psi_u + \Psi_v + \frac{h}{\sqrt{g h}} \sqrt{\Psi^2+1} \Psi_h = 0,\\
&-(\sqrt{3} + \Psi) \Psi_u + (\sqrt{3} \Psi - 1) \Psi_v + \frac{2 h}{\sqrt{g h}} \sqrt{1+\Psi^2} \Psi_h = 0,
\end{split}
\end{equation}
implying that $\Psi$ must satisfy
\begin{equation}
\label{2-2-Psi-compat}
(1+\sqrt{3} \Psi) \Psi_u + (\sqrt{3} - \Psi) \Psi_v = 0.
\end{equation}
It is easy to show from (\ref{2-2-system-for-Psi}) that $\Phi$ is either constant or depends essentially on all functions $u,v,h$. Looking for a solution of the form $\Psi = F(\gamma_1 (\Psi) u + \gamma_2(\Psi) v - \phi(h))$, where $\phi(h)$ is some function of $h$ to be determined, we obtain that equations (\ref{2-2-system-for-Psi}) possess the implicit solution
\begin{equation}
\label{EE-non-constant-Psi-1}
\Psi = F\left((\Psi - \sqrt{3}) u + (1+\sqrt{3} \Psi) v - 2\sqrt{1+\Psi^2} \sqrt{g h} \right),
\end{equation}
where $F$ is an arbitrary function of its argument.  Equations (\ref{2-2-system-for-Psi}) also possess infinite classes of solutions of the form
\begin{equation}
\label{E-E-equation-for-Psi}
\Psi = F \left( s_1, s_2 \right), \quad s_1 = \gamma_1(\Psi) u - 2\sqrt{g h}, \quad s_2 = \gamma_2(\Psi) v - 2\sqrt{g h}.
\end{equation}
The compatibility relation (\ref{2-2-Psi-compat}) requires that
\begin{equation}
\label{2-2-2-equation-for-F}
\pd{F}{s_1} = \frac{(\Psi - \sqrt{3}) \gamma_2(\Psi)}{(1+\sqrt{3} \Psi) \gamma_1(\Psi)} \pd{F}{s_2} = G(s_1,s_2) \pd{F}{s_2},
\end{equation}
for some function $G(s_1,s_2)$.  Equations (\ref{2-2-system-for-Psi}) then become
\begin{equation}
\label{2-2-gamma12}
\gamma_1(\Psi) (\sqrt{3} \Psi^2 - 2\Psi - \sqrt{3}) + \gamma_2(\Psi) (\Psi^2 - 2\sqrt{3} \Psi + 3) + \gamma_1(\Psi)\gamma_2(\Psi) \sqrt{1+\Psi^2}(\sqrt{3}-\Psi) = 0.
\end{equation}
For a selected function $G(s_1,s_2)$, solving equations (\ref{2-2-gamma12}) and
$$G(s_1,s_2) = \frac{(\Psi - \sqrt{3}) \gamma_2(\Psi)}{(1+\sqrt{3} \Psi) \gamma_1(\Psi)}$$
gives the explicit expressions for $\gamma_1(\Psi)$ and $\gamma_2(\Psi)$ while integration of (\ref{2-2-2-equation-for-F}) gives the dependence of $F$ on $s_1$ and $s_2$.
For example, when $G(s_1,s_2) = 1$, then $\Psi = F(\gamma_1(\Psi) + \gamma_2(\Psi) - 4\sqrt{g h} )$, with
\begin{equation}
\gamma_1(\Psi) = 2 \frac{\sqrt{3} \Psi^3 - 5 \Psi^2 + \sqrt{3} \Psi + 3}{(\sqrt{3} \Psi^2 - 2\Psi - \sqrt{3}) \sqrt{1+\Psi^2}}, \quad 
\gamma_2(\Psi) = 2 \frac{3 \Psi^4 - 4\sqrt{3}\Psi^3 - 2\Psi^2 + 4\sqrt{3} \Psi + 3}{(\sqrt{3} \Psi^3 - 5 \Psi^2 + \sqrt{3} \Psi + 3)\sqrt{1+\Psi^2}},
\end{equation}
and $F$ arbitrary.

From any explicit solution of (\ref{2-2-traces-eqs-c}) obtained by specifying the arbitrary function in (\ref{EE-non-constant-Psi-1}) or in (\ref{E-E-equation-for-Psi}) and (\ref{2-2-2-equation-for-F}) and using the relations (\ref{EE-nc-1112}), (\ref{EE-nc-2122}), the solution for the vector fields $u(r^1,r^2), v(r^1,r^2)$ is obtained by integrating system (\ref{EE-solved}).  However, since the resulting expressions are very involved even in the simplest cases, we will not present a solution of this type in closed form.

{\bf iv) } Finally, conducting an analysis similar to that of the previous case, we finally look for linear interactions of two acoustic-type waves of constant direction for which we choose different signs for $\varepsilon$ in (\ref{wave-vectors} ii).  Suppose in this case that the Riemann invariants are given in the form
\begin{equation}
\begin{split}
&r^1 = -(\lambda^1_1 u + \lambda^1_2 v + \sqrt{g h})t + \lambda^1_1 x + \lambda^1_2 y, \quad \lambda^i_j \in \mathbb{R},\\
&r^2 = -(\lambda^2_1 u + \lambda^2_2 v - \sqrt{g h})t + \lambda^2_1 x + \lambda^2_2 y, \quad |\vec{\lambda}^i| = 1, i=1,2.
\end{split}
\end{equation}
Writing $\lambda^1_1 = \sin{\varphi_1}, \lambda^1_2 = \cos{\varphi_1}, \lambda^2_1 = \sin{\varphi_2}, \lambda^2_2 = \cos{\varphi_2}$, where $\varphi_1$, $\varphi_2$ are constant, we find that a rank-2 solution invariant under
\begin{equation}
\begin{split}
X &= \sin{(\varphi_1 - \varphi_2)} \p_t + \left( \sin{(\varphi_1 - \varphi_2)} u + (\cos{(\varphi_1)}  + \cos{(\varphi_2)})\sqrt{g h}\right)\p_x\\
&\qquad\quad\quad + \left( \sin{(\varphi_1 - \varphi_2)} v - (\sin{(\varphi_1)} + \sin{(\varphi_2)})\sqrt{g h}\right)\p_y
\end{split}
\end{equation}
exists if and only if the angle between $\varphi_1$ and $\varphi_2$ satisfies
\begin{equation}
\label{E-E-2-angle-var}
|\varphi_1 - \varphi_2| = \frac{\pi}{3}, 
\end{equation}
in comparison with relation (\ref{E-E-angle-var}).  This nonscattering rank-2 solution of the SWW equations can be presented as
\begin{equation*}
\begin{split}
&u = u_0 + 2\sqrt{g} \left(\lambda^1_1 h_1(r^1) - \lambda^2_1 h_2(r^2)\right), \quad v = v_0 + 2\sqrt{g} \left(\lambda^1_2 h_1(r^1) - \lambda^2_2 h_2(r^2)\right),\\
&h = \left(h_1(r^1) + h_2(r^2)\right)^2, \quad u_0, v_0 \in\mathbb{R},
\end{split}
\end{equation*}
where the functions $h_1(r^1)$ and $h_2(r^2)$ are arbitrary functions of the Riemann invariants
\begin{equation}
\begin{split}
&r^1 = -\left(\lambda^1_1 u_0  + \lambda^1_2 v_0 + 3\sqrt{g} h_1(r^1) \right) t + \lambda^1_1 x + \lambda^1_2 y, \quad \lambda^i_j \in \mathbb{R}, \quad \vec{\lambda}^1 \cdot \vec{\lambda}^2 = 1/2,\\
&r^2 = -\left(\lambda^2_1 u_0 + \lambda^2_2 v_0  - 3\sqrt{g} h_2(r^2) \right) t + \lambda^2_1 x + \lambda^2_2 y, \quad |\vec{\lambda}^i| = 1, \quad i=1,2.
\end{split}
\end{equation}
so that the angle between $\vec{\lambda}^1$ and $\vec{\lambda}^2$ satisfies (\ref{E-E-2-angle-var}).  The similarity with solution (\ref{EE-sol-nonscattering}) is not surprising.  It can in fact be obtained by considering the wave vector $\vec{\lambda}^2$ in the opposite direction, that is by setting $\vec{\lambda}^2 \to -\vec{\lambda}^2$ in expressions (\ref{E-E-angle-var}), (\ref{EE-sol-nonscattering}) and (\ref{EE-RI-nonscattering}).  The computation of the corresponding solution of the RSWW equations is done analogously to that of the previous case and the result is included in Table \ref{Table-Rank-2-RSWW}.

\section{Conclusion}
In this work, we have extended the applicability of the conditional symmetry approach in the context of Riemann invariants to a certain class of first order inhomogeneous quasilinear hyperbolic system of the first order, namely those systems that are equivalent to a homogeneous one under an invertible point transformation.  Such classes of systems have been characterized recently in the case of systems of two equations in two dependent and independent variables in \cite{Curro-Oliveri-JMP-2008} and an algorithm to construct the appropriate point transformation was also given. The key element in this analysis is the presence of an infinite dimensional Lie algebra admitted by every quasilinear homogenous system in two variables.  Although this is not true in general for multidimensional systems, we have been able to show that such a transformation exists for the rotating shallow water wave equations and after an analysis of the rank-$k$ solutions of the SWW equations, we used it to construct several of their implicit solutions expressed in terms of Riemann invariants.  While several classes of invariant solutions of the RSWW equations are known, these new conditionally invariant solutions possess in general a considerable degree of freedom in the sense that they depend on one or two arbitrary functions of the Riemann invariants.  Although it is possible in the case of a homogeneous system to select these arbitrary functions so as to obtain bounded solutions for every value of the Riemann invariants,
such solutions could not be constructed here since the point transformation (\ref{SWW-RSWW-iso}) is singular at times $t=\frac{\pi}{2\Omega} (2 n + 1), n \in \mathbb{N}$.  However, by using invariance under time translation, we have shown that it is possible to construct solutions expressed in terms of Riemann invariants defined in a finite interval around $t=0$.

One may ask whether rank-$k$ solutions of a given inhomogeneous system in the form (\ref{rank-k-solution}) can be constructed without relying on a point transformation bringing it to a homogeneous form.  A preliminary analysis shows that this type of solution would possess invariance properties similar to those admitted by homogeneous systems, as expressed in Proposition 1.  This study shall be addressed in a future work.

{\bf Acknowledgement : } This work has been supported by a research fellowship from NSERC of Canada. The author thanks Professor A.M. Grundland (Centre de Recherches Math\'ematiques at the Universit\'e de Montr\'eal and Universit\'e du Qu\'ebec \`a Trois-Rivi\`eres) for helpful and interesting discussions on the topic of this paper.

\begin{landscape}
\vspace{-5cm}
\begin{table}
\begin{math}
\begin{array}{|l|l|l|l|}
\hline
\text{Type} & \text{Solution} & \text{Riemann invariant} & \text{Comments} \\\hline
1.E	& u = u_0 - \frac{\lambda_2}{\lambda_1}\varphi(r)	& r = - u_0 \lambda_1 t + \lambda_1 x + \lambda_2 y 						& \varphi : \mathbb{R}\to\mathbb{R}\\
	& v = \varphi(r)		 			&	 											& \lambda_i, u_0 \in \mathbb{R}\\
	& h = h_0 						& 												& h_0 \in \mathbb{R}^+\\ 
\hline
2.E	& u = C \sin{r}						& r = C ( -C t + x \sin{r} + y \cos{r})								& C \in \mathbb{R}\\
	& v = C \cos{r}						& 												& h_0 \in \mathbb{R}^+\\
	& h = h_0						&												& \\
\hline
3.E & u = \varphi(r)					& r = -2 C t + \frac{C}{\varphi(r)} x + u(r) y							& \varphi : \mathbb{R}\to\mathbb{R}\\
	& v = C/\varphi(r)					& 												& C \in \mathbb{R} \\
	& h = h_0						&												& h_0 \in \mathbb{R}^+\\
\hline
4.S	& u = u_0 + 2\lambda_1 \sqrt{g} \varphi(r)		& r = -(\lambda_1 u_0 + \lambda_2 v_0 + 3 \sqrt{g h})t + \lambda_1 x + \lambda_2 y		& \varphi : \mathbb{R}\to\mathbb{R}\\
	& v = v_0 + 2\lambda_2 \sqrt{g} \varphi(r)		& 												& \lambda_i, u_0,v_0 \in \mathbb{R}\\
	& h = \varphi(r)^2					&												& \\
\hline
5.S	& u = u_0 - 2 \sqrt{g} \cos{\varphi(r)}		& r = -\left(u_0 \sin(\varphi(r)) + v_0 \cos(\varphi(r))+\sqrt{g}(\varphi(r)+h_0)\right)t 		& \varphi : \mathbb{R}\to\mathbb{R} \\
	& v = v_0 + 2 \sqrt{g} \sin{\varphi(r)}		& \qquad +\sin(\varphi(r)) x + \cos(\varphi(r)) y											    & u_0, v_0 \in \mathbb{R} \\
	& h = (\varphi(r)+h_0)^2					        &												                                                & h_0 \in \mathbb{R}^+\\
\hline
6.S	& u = u_0 + \sqrt{2\pi g} S\left(\sqrt{\frac{2 \varphi(r)}{\pi}}\right)	& r = -\Bigg(\sin(\varphi(r))\left(u_0 + \sqrt{2\pi g} S\left(\sqrt{\frac{2 \varphi(r)}{\pi}}\right)\right) & \varphi : \mathbb{R}\to\mathbb{R}^+\\
	& v = v_0 + \sqrt{2\pi g} C\left(\sqrt{\frac{2 \varphi(r)}{\pi}}\right)	& \quad \qquad +\cos(\varphi(r))\left(v_0 + \sqrt{2\pi g} C\left(\sqrt{\frac{2 \varphi(r)}{\pi}}\right)\right) & u_0,v_0 \in \mathbb{R} \\
	& h = (\sqrt{\varphi(r)}+h_0)^2        			& \quad \qquad + \sqrt{g}(\sqrt{\varphi(r)}+h_0) \Bigg)t + \sin(\varphi(r)) x + \cos(\varphi(r)) y			& h_0 \in \mathbb{R}^+\\
\hline
\end{array}
\end{math}
\caption{Rank-1 solutions of the SWW equation (\ref{SWW-eqs}).  The functions $S(\cdot)$ and $C(\cdot)$ are the sine and cosine Fresnel integrals.}
\label{Table-rank-1-solutions}
\end{table}
\end{landscape}

\begin{landscape}

\begin{table}
\begin{math}
\begin{array}{|l|l|l|l|} 
\hline
Type 	& \text{Riemann invariants} & \text{Solution} & \text{Comments} \\\hline

ES 	& r^1 = \lambda^1_2(v) \left( \left(2 G(s) + \sqrt{3} \varepsilon F(r^2)\right)t - \sqrt{3} \varepsilon x - y\right)	& u = \frac{\sqrt{3}}{3} \varepsilon G(s) + F(r^2) 	& h_0 \in \mathbb{R}^+, \varepsilon^2=1\\
    	& r^2 = \left(\frac{3}{2} F(r^2) + \frac{h_0}{2}\right) t - x								& v = G\left(s\right)					& F,G : \mathbb{R} \to \mathbb{R}\\
	& s = r^2 - \frac{\sqrt{3} \varepsilon}{2 \lambda^1_2 (G(s))} r^1								& h = \frac{1}{4 g} \left( F(r^2) - \frac{2\sqrt{3}}{3} \varepsilon G(s) + h_0 \right)^2	&  \lambda^1_2 : \mathbb{R} \to \mathbb{R}\\\hline

SS	& r^1 = - \left(\lambda^1_1 u_0 + \lambda^1_2 v_0 + 3 \sqrt{g} h_1(r^1)\right) t + \lambda^1_1 x + \lambda^1_2 y	& u = u_0 + 2\sqrt{g} \left(\lambda^1_1 h_1(r^1) + \varepsilon \lambda^2_1 h_2(r^2)\right)	&  u_0, v_0, \lambda^i_j \in \mathbb{R}, |\vec{\lambda}^i| = 1\\
	& r^2 = - \left(\lambda^2_1 u_0 + \lambda^2_2 v_0 + 3\varepsilon \sqrt{g} h_2(r^2)\right) t + \lambda^2_1 x + \lambda^2_2 y	& v = v_0 + 2\sqrt{g} \left(\lambda^1_2 h_1(r^1) + \varepsilon \lambda^2_2 h_2(r^2)\right)	& h_1, h_2 : \mathbb{R}\to\mathbb{R}\\
	& 															& h = \left(h_1(r^1) + h_2(r^2)\right)^2	&  \vec{\lambda}^1 \cdot \vec{\lambda}^2 = -\varepsilon/2, \varepsilon^2=1\\
\hline

\end{array}
\end{math}
\caption{Rank-2 solutions of the SWW equations.}
\label{Table-Rank-2-SWW}
\end{table}

\begin{table}
\begin{math}
\begin{array}{|l|l|l|l|} 
\hline

Type 	& \text{Riemann invariants} & \text{Solution} & \text{Comments} \\\hline
ES	& r^1 = \lambda^1_2\left(G(\tilde{s})\right) \Big( -\frac{1}{2\Omega} (2 G(s) + \sqrt{3} \varepsilon F(r^2)) \cot{(\Omega t)} & u = {\scriptstyle - \left( \frac{\sqrt{3}}{3} \varepsilon G(s) + F(r^2) \right) \cot{\left(\Omega t\right)} - G(s) + \Omega (y+x \cot{(\Omega t)}) }  &   h_0 \in \mathbb{R}^+, \epsilon^2 =1     \\ 
	& \quad - \frac{\sqrt{3}}{2} \varepsilon (y - x \cot{(\Omega t)}) + \frac{1}{2} (x + y \cot{(\Omega t)}) \Big)								& v = {\scriptstyle \frac{\sqrt{3}}{3} \varepsilon G(s) + F(r^2) - G(s) \cot{(\Omega t)} - \Omega (x - y \cot{(\Omega t)}) }	&  F, G : \mathbb{R}\to\mathbb{R}\\	
	& r^2 = - \frac{1}{4\Omega} \left( 3 F(r^2) + h_0 \right) \cot{(\Omega t)} - \frac{1}{2} (y - x \cot{(\Omega t)}), & h = {\scriptstyle \frac{1}{4 g} \left( F(r^2) - \frac{2 \sqrt{3}}{3} \varepsilon G(s) + h_0 \right)^2 \csc^2{(\Omega t)} } & \lambda^1_2 : \mathbb{R} \to \mathbb{R}\\
	& s = r^2 - \frac{\sqrt{3} \varepsilon}{2 \lambda^1_2 \left(G(s)\right)} r^1  & & \\ 
\hline

SS  & r^1 = \frac{1}{2} \Big[ \frac{1}{\Omega}(\lambda^1_1 u_0 + \lambda^1_2 v_0 + 3 \sqrt{g} h_1(r^1)) \cot{(\Omega t)} & u = { \scriptstyle -(u_0 +2 \sqrt{g}(\lambda^1_1 h_1({r}^1) + \varepsilon \lambda^2_1 h_2({r}^2))) \cot{(\Omega t)} , } & u_0, v_0, \lambda^i_j \in \mathbb{R}, \\
    & \quad + \lambda^1_1 (y- x \cot{(\Omega t)}) - \lambda^1_2 (x + y \cot{(\Omega t)})\Big] &  {\scriptstyle \quad - (v_0 + 2 \sqrt{g} (\lambda^1_2 h_1({r}^1) + \varepsilon \lambda^2_2 h_2({r}^2))) + \Omega (y+x \cot{(\Omega t)}) } & h_1, h_2 : \mathbb{R}\to\mathbb{R}\\
    & r^2 = \frac{1}{2} \Big[ \frac{1}{\Omega}(\lambda^2_1 u_0 + \lambda^2_2 v_0 + 3\varepsilon \sqrt{g} h_2(r^2)) \cot{(\Omega t)} &  v = {\scriptstyle - (v_0 + 2 \sqrt{g} (\lambda^1_2 h_1({r}^1) +\varepsilon \lambda^2_2 h_2({r}^2))) \cot{(\Omega t)}     } & \vec{\lambda}^1 \cdot \vec{\lambda}^2 = -\varepsilon/2\\
    &\quad + \lambda^2_1 (y- x \cot{(\Omega t)}) - \lambda^2_2 (x + y \cot{(\Omega t)})\Big] & {\scriptstyle \quad + u_0 + 2 \sqrt{g}(\lambda^1_1 h_1({r}^1) + \varepsilon \lambda^2_1 h_2({r}^2)) - \Omega (x - y \cot{(\Omega t)})} & |\vec{\lambda}^i|=1, \,\varepsilon^2 = 1 \\
    &                                                                                                           & h = {\scriptstyle (h_1(\tilde{r}^1) + h_2(\tilde{r}^2)) \csc^2({\Omega t}) }&  \\\hline 

\end{array}
\end{math}
\caption{Rank-2 solutions of the RSWW equations.}
\label{Table-Rank-2-RSWW}
\end{table}

\begin{table}
\hspace{-1cm}
\begin{math}
\begin{array}{|l|l|l|l|}
\hline
\text{No}       & \text{Riemann invariants}                                         & \text{Solution}                                                                              & \text{Comments}\\ \hline
1.              & r^1 = (2 \tanh^2{(s)} + \sqrt{3} \tanh^2{(r^2)})t - \sqrt{3}x -y      &   u = \frac{\sqrt{3}}{3} \tanh^2{(s)} + \tanh^2{(r^2)}                                           & \text{Anti-bump} \\
                & r^2 = \left(\frac{3}{2} \tanh^2{(r^2)} + \frac{h_0}{2}\right)t - x  &   v = \tanh^2{(s)}                                                                              & \\
                & s = r^2 - \frac{\sqrt{3}}{2} r^1                                   &   h = \frac{1}{4g} \left(\tanh^2{(r^2)} - \frac{2\sqrt{3}}{3} \tanh^2{(s)} + h_0\right)^2         & \\ 
\hline
2.              & r^1 = (2 \sech^2{(s)} + \sqrt{3} \sech^2{(r^2)})t - \sqrt{3}x -y      &   u = \frac{\sqrt{3}}{3} \sech^2{(s)} + \sech^2{(r^2)}                                           & \text{Bump} \\
                & r^2 = \left(\frac{3}{2} \sech^2{(r^2)} + \frac{h_0}{2}\right)t - x  &   v = \sech^2{(s)}                                                                              & \\
                & s = r^2 - \frac{\sqrt{3}}{2} r^1                                   &   h = \frac{1}{4g} \left(\sech^2{(r^2)} - \frac{2\sqrt{3}}{3} \sech^2{(s)} + h_0\right)^2         & \\ 
\hline

3.              & r^1 = -\left( u_0 + 3 \sqrt{g} \sech^2(r^1) \right) t + x                              
                & u = u_0 + 2\sqrt{g} \left( \sech^2(r^1) - \frac{1}{2} \sech^2(r^2)  \right)   &  \text{Bump} \\
                & r^2 = -\left(-\frac{u_0}{2} + \frac{\sqrt{3}}{2} v_0 + 3 \sqrt{g} \sech^2(r^2) \right) t - \frac{1}{2} x + \frac{\sqrt{3}}{2} y   
		& v = v_0 + \sqrt{3 g} \sech^2(r^2) & u_0,v_0Â \in \mathbb{R}\\
                &                           &  h = \left( \sech^2(r^1) + \sech^2(r^2)  \right)^2  & \\
\hline          

4.              & r^1 = -\left(u_0 + \frac{3 \sqrt{g} A_1 r^1}{\sqrt{1 + B_1 (r^1)^2 }} \right) t + x                              
                & u = u_0 + 2\sqrt{g} \left( \frac{A_1 r^1}{\sqrt{1+B_1 (r^1)^2}} - \frac{A_2 r^2}{2 \sqrt{1 + B_2 (r^2)^2}} \right)   &  \text{Kink} \\
                & r^2 = -\left(-\frac{u_0}{2} + \frac{\sqrt{3}}{2} v_0 + \frac{3 \sqrt{g} A_2}{\sqrt{1+B_2 (r^2)^2}} \right) t - \frac{1}{2} x + \frac{\sqrt{3}}{2} y   
		& v = v_0 + \frac{\sqrt{3 g} A_2}{ \sqrt{1 + B_2 (r^2)^2}  } & u_0,v_0, A_1, A_2Â \in \mathbb{R}\\
                & 				
		&h = \left( \frac{A_1 r^1}{\sqrt{1+B_1 (r^1)^2}} + \frac{A_2 r^2}{ \sqrt{1 + B_2 (r^2)^2}} \right)^2  & B_1, B_2 \in \mathbb{R}^{+}\\
\hline

5.              & r^1 = -\left(u_0 + \frac{3 \sqrt{g} A_1 }{ \wp{\left(r^1, \frac{4}{3}, \frac{8}{27} + \frac{4}{3} A_1^4\right)}    } \right) t + x                              
                & u = u_0 + 2\sqrt{g} \left( \frac{A_1 }{ \wp{\left(r^1, \frac{4}{3}, \frac{8}{27} + \frac{4}{3} A_1^4\right)}} - \frac{1}{2} \frac{A_2 }{ \wp{\left(r^2, \frac{4}{3}, \frac{8}{27} + \frac{4}{3} A_2^4\right)} } \right)  &  \text{Periodic} \\
                & r^2 = -\left(-\frac{u_0}{2} + \frac{\sqrt{3}}{2} v_0 + \frac{A_2 }{ \wp{\left(r^2, \frac{4}{3}, \frac{8}{27} + \frac{4}{3} A_2^4\right)}} \right)   t - \frac{1}{2} x + \frac{\sqrt{3}}{2} y   
		& v = v_0 + \frac{A_2 }{ \wp{\left(r^2, \frac{4}{3}, \frac{8}{27} + \frac{4}{3} A_2^4\right)}} & u_0,v_0, A_1, A_2Â \in \mathbb{R}\\
                & 				
		&h = \left( \frac{A_1 }{ \wp{\left(r^1, \frac{4}{3}, \frac{8}{27} + \frac{4}{3} A_1^4\right)}} + \frac{A_2 }{ \wp{\left(r^2, \frac{4}{3}, \frac{8}{27} + \frac{4}{3} A_2^4\right)}}    \right)^2  & \\
\hline
\end{array}
\end{math}
\caption{Examples of bounded rank-2 solutions of the SWW equations. The function $\wp(\cdot,g_2,g_3)$ is the elliptic Weierstrass $\wp$ function with invariants $g_2$,$g_3$. }
\label{Table-Rank-2-Bounded}
\end{table}

\end{landscape}

\bibliographystyle{plain}
\bibliography{Biblio}

\begin{thebibliography}{10}

\bibitem{Hereman-symmgrp}
B.~Champagne, W.~Hereman, and P.~Winternitz.
\newblock The computer calculation of {L}ie point symmetries of large systems
  of differential equations.
\newblock {\em Comput. Phys. Comm.}, 66(2-3):319--340, 1991.

\bibitem{Chesnokov-SWW}
A.A. Chesnokov.
\newblock Symmetries and exact solutions of shallow water equations on a
  three-dimensional shear flow.
\newblock {\em Prikl. Mekh. Tekhn. Fiz.}, 49(5):41--54, 2008.

\bibitem{Chesnokov-RSWW}
A.A. Chesnokov.
\newblock Symmetries and exact solutions of the rotating shallow water
  equations.
\newblock {\em Europ. J. Appl. Math.}, 20:461--477, 2009.

\bibitem{Conte-Grundland-Huard-JPhysA-2009}
R.~Conte, Grundland A.M., and B.~Huard.
\newblock Elliptic solutions of isentropic ideal compressible fluid flow in
  (3+1) dimensions.
\newblock {\em Journal of Physics A : Mathematical and Theoretical}, 42(13),
  2009.

\bibitem{Curro-Oliveri-JMP-2008}
C.~Curr\`o and F.~Oliveri.
\newblock Reduction of nonhomogeneous quasilinear 2{$\times$}2 systems to
  homogeneous and autonomous form.
\newblock {\em Journal of Mathematical Physics}, 49(10):103504, 2008.

\bibitem{Donato-Oliveri-MHD-1993}
A.~Donato and F.~Oliveri.
\newblock Reduction to autonomous form by group analysis and exact solutions of
  axisymmetric {MHD} equations.
\newblock {\em Math. Comput. Modelling}, 18(10):83--90, 1993.
\newblock Similarity, symmetry and solutions of nonlinear boundary value
  problems (Wollongong, 1992).

\bibitem{Donato-Oliveri-1995}
A.~Donato and F.~Oliveri.
\newblock When nonautonomous equations are equivalent to autonomous ones.
\newblock {\em Appl. Anal.}, 58(3-4):313--323, 1995.

\bibitem{Grundland-Huard-JNMP}
A.M. Grundland and B.~Huard.
\newblock {R}iemann invariants and rank-k solutions of hyperbolic systems.
\newblock {\em Journal of Nonlinear Mathematical Physics}, 13(3):393--419,
  2006.

\bibitem{Grundland07-1}
A.M. Grundland and B.~Huard.
\newblock Conditional symmetries and {R}iemann invariants for hyperbolic
  systems of pdes.
\newblock {\em Journal of Physics A Mathematical and Theoretical}, 40(15):4093,
  2007.

\bibitem{Hereman-review-94}
W.~Hereman.
\newblock Review of symbolic software for the computation of lie symmetries of
  differential equations.
\newblock {\em Euromath Bull}, 1:45--79, 1994.

\bibitem{Hereman-symmgrp2009}
W.~Hereman and B.~Huard.
\newblock symmgrp2009.max : A macsyma/maxima program for the calculation of lie
  point symmetries of large systems of differential equations.
\newblock \url{http://inside.mines.edu/~whereman/}.

\bibitem{Levi-Winternitz-Nucci-Rogers-1989}
D.~Levi, M.~C. Nucci, C.~Rogers, and P.~Winternitz.
\newblock Group theoretical analysis of a rotating shallow liquid in a rigid
  container.
\newblock {\em J. Phys. A}, 22(22):4743--4767, 1989.

\bibitem{Olver-2000}
P.J. Olver.
\newblock {\em Applications of Lie Groups to Differential Equations}.
\newblock Springer, New York, 2000.

\bibitem{Peradzynski-GMC}
Z.~Peradzynski.
\newblock On certain classes of exact solutions for gasdynamics equations.
\newblock {\em Arch. Mech.}, 24:287--303, 1972.

\end{thebibliography}

\appendix

\end{document}